\documentclass[12pt,preprint]{aastex}
\def\E{\varepsilon}

\shorttitle{$^3$He and $^4$He Acceleration}
\shortauthors{Liu et al.}

\begin{document}


\title{Stochastic Acceleration of $^3$He and $^4$He in Solar Flares by Parallel Propagating 
Plasma Waves: General Results}
\vspace{1cm}

\author{Siming Liu\altaffilmark{1}, Vah\'{e} Petrosian\altaffilmark{1, 2}, and Glenn M. 
Mason\altaffilmark{3}}


\altaffiltext{1}{Center for Space Science and Astrophysics, Department of Physics, Stanford
University, Stanford, CA 94305; liusm@stanford.edu}
\altaffiltext{2}{Department of Physics and Applied Physics, Stanford University, Stanford, 
CA 94305; vahe@astronomy.stanford.edu}
\altaffiltext{3}{Department of Physics and Institute for Physical Sciences and Technology, 
University of Maryland, College Park, MD 20742-4111; glenn.mason@umail.umd.edu}


\begin{abstract}

In this work we study the acceleration in solar flares of $^3$He and $^4$He from a thermal background 
by parallel propagating plasma waves with a general broken power-law spectrum that takes into account 
the turbulence generation processes at large scales and the thermal damping effects at small scales. 
The exact dispersion relation for a cold plasma is used to describe the relevant wave modes. In the 
nonrelativistic regime of interest here the charged particle acceleration is dominated by resonant 
interactions with high frequency waves, and the pitch angle diffusion rate is usually smaller than 
the momentum diffusion rate below a few MeV nucleon$^{-1}$. In the absence of other more efficient 
scattering agents, this may give rise to an anisotropic particle distribution. Because low-energy 
$\alpha$-particles only interact with small scale waves in the $^4$He-cyclotron branch, where the 
wave frequencies are below the $\alpha$-particle gyro-frequency, their pitch angle averaged 
acceleration time is at least one order of magnitude longer than that of $^3$He ions, which mostly 
resonate with relatively higher frequency waves in the proton-cyclotron (PC) branch. The 
$\alpha$-particle acceleration rate starts to approach that of $^3$He beyond a few tens of keV 
nucleon$^{-1}$, where $\alpha$-particles can also interact with long wavelength waves in the PC 
branch. However, the $^4$He acceleration rate is always smaller than that of $^3$He. Consequently, 
the acceleration of $^4$He is suppressed significantly at low energies, and the spectrum of the 
accelerated $\alpha$-particles is always softer than that of $^3$He. The effects of thermal damping 
further magnify this suppression. On the other hand, the length scale of waves in the PC branch at 
generation (the longest scale) mostly affects the high-energy cutoffs of the accelerated particle 
spectra. These favorable conditions for $^3$He acceleration diminish in strongly magnetized plasmas 
and/or for strong turbulence. The model gives reasonable account of the observed low-energy $^3$He 
and $^4$He fluxes and spectra in the impulsive solar energetic particle events observed with the {\it 
Advanced Composition Explorer}. Other acceleration processes and/or stochastic acceleration by other 
wave modes seem to be required to explain the occasionally observed decrease of $^3$He to $^4$He 
ratio at energies beyond a few MeV nucleon$^{-1}$.

\end{abstract}



\keywords{acceleration of particles --- plasma --- Sun: abundances --- Sun: 
flares --- turbulence}


\section{INTRODUCTION}

Stochastic acceleration (SA) of particles by plasma waves or turbulence (PWT), a second order Fermi 
acceleration process, plays an important role in understanding the energy release processes and the 
consequent plasma heating and particle acceleration in solar flares (e.g., Ramaty 1979; M\"{o}bius et 
al. 1980, 1982;  Miller et al. 1997; Petrosian \& Liu 2004, hereafter PL04). This theory was applied 
to the acceleration of nonthermal electrons (Miller \& Ramaty 1987;  Hamilton \& Petrosian 1992; 
Park, Petrosian \& Schwartz 1997;  Petrosian \& Donaghy 1999), which are responsible for the 
microwave and hard X-ray emissions and for type III radio bursts during the impulsive phase of solar 
flares. In this respect it has achieved some degree of success (see PL04). It has also been advocated 
for production of nonthermal electrons and ions observed near the Earth in association with solar 
flares (Reames, von Rosenvinge \& Lin 1985; Mason et al.  1989; Van Hollebeke, McDonald \& Meyer 
1990; Bieber et al. 1994). The accelerated ions show charge-to-mass ratio dependent enhancements 
relative to the photospheric values, which are readily explained in the context of SA (Miller 2003; 
Hurford et al. 1975; Reames, Meyer \& von Rosenvinge 1994; Mason et al. 1986, 2002, 2004; Reames \& 
Ng 2004). In a few gamma-ray flares, 
where gamma-ray line emissions due to ion nuclear interactions are clearly observed, there is also 
evidence for an anomalous abundance pattern of the accelerated ions (Share \& Murphy 1998; Hua, 
Ramaty \& Lingenfelter 1989). The competing models of diffusive shock acceleration and/or direct 
acceleration by parallel (to magnetic fields) electric fields have not been subjected to quantitative 
tests. In addition maintaining large electric fields appears to be problematical (see, however, 
Holman 1985), and there is no convincing evidence for the presence of low coronal shocks for the 
impulsive flares.

The most critical challenge to these models including the SA model arises from the extreme 
enhancement of $^3$He observed in some scatter free impulsive events (Hsieh \& Simpson 1970; 
Serlemitsos \& Balasubrahmanyan 1975; Mason, Dwyer \& Mazur 2000). All models 
proposed for this are based on either resonant plasma heating or resonant particle acceleration (Fisk 
1978; Temerin \& Roth 1992; Miller \& Vi\~{n}as 1993; Zhang 1999; Paesold, Kallenbach \& Benz 2003). 
We recently carried out a quantitative study and showed that the SA of $^3$He and $^4$He by parallel 
propagating plasma waves can account for the $^3$He enhancement, its varied range, and the spectral 
shape (Liu, Petrosian \& Mason 2004, hereafter LPM04). The primary feature of our new finding, which 
comes from our use of the exact dispersion relation, is that the acceleration of $^4$He from a 
low-energy thermal background is suppressed due to its lack of resonance with multiple-waves, a 
result very similar to the suppression of proton acceleration relative to the electron acceleration 
(Schlickeiser 1989; PL04).

Earlier studies of SA of low-energy thermal background electrons and protons revealed an acceleration 
rate larger than the scattering rate (Pryadko \& Petrosian 1997;  PL04). We show here that, for 
coronal conditions, this is also true for the SA of $^3$He and $^4$He. Consequently, in impulsive 
events SA by PWT at low energies (determined by the acceleration rate) is more efficient than the 
acceleration by diffusive shocks, whose rate of acceleration is proportional to the scattering rate. 
In addition, recent observations of gradual events, which may involve acceleration by interplanetary 
shocks, suggest that the source population is pre-accelerated by a distinct acceleration process 
operating in impulsive or gradual flares (Mason, Mazur \& Dwyer 1999; Desai et al. 2001, 2003, 2004). 
SA may very well be the agent for this acceleration as well.

In this paper we explore more realistic models and quantify the dependence of $^3$He enhancement on 
basic model parameters. In the next section we discuss the basic features of resonant wave-particle 
interaction. After showing that the scattering rate is usually lower than the acceleration rate at 
low energies, we calculate the
pitch angle averaged acceleration timescales of $^3$He and $^4$He under different plasma
conditions and for different turbulence spectra, taking into account the uncertainties
associated with the generation of turbulence, its cascade and damping. Surprisingly, the
acceleration time of $^3$He is always shorter than that of $^4$He at nonrelativistic energies.  
In \S\ \ref{sep} the model of SA is briefly summarized and applied to observations of six
impulsive solar energetic particle events (SEPs) observed with the {\it Advanced Composition
Explorer} ({\it ACE}), where the low-energy ion spectra have convex shapes, which are referred
to as \underline{rounded spectra} (Mason et al. 2002). Reasonable fits to the observed $^3$He
and $^4$He spectra are obtained by adjusting the temperature of the background plasma, the
turbulence generation length scale, the intensity of the turbulence, and a normalization factor.
In \S\ \ref{dis} we explore the model parameter space to see how observations can be used to
constrain properties of the turbulence and the flaring plasma.  It is shown that the
acceleration of low-energy $^4$He can still be suppressed, even without the thermal damping
effects, because $^4$He interacts mostly with small scale waves that have relatively low phase
velocities. The thermal damping effects just amplify this suppression. The generation length
scale of waves in the proton cyclotron (PC) branch determines the acceleration rates at high
energies and thus mostly affects the high-energy cutoffs. On the other hand, we
show that in more strongly magnetized plasmas the accelerations of $^4$He and $^3$He become
comparable, and $^3$He enhancement decreases. This is also true with the increase of the
background plasma temperature and/or the intensity of the turbulence. 
In \S\ \ref{sum}, we summarize the
main results and emphasize the importance of studying the properties of the PC branch in
understanding the relative acceleration of $^3$He and $^4$He. The model limitation and future
developments of the SA theory are also discussed.

\section{RESONANT WAVE-PARTICLE INTERACTION}
\label{int}


SA of particles by parallel propagating waves has been explored in PL04. For the sake of 
completeness, we summarize here the main results relevant to current study.

\subsection{Dispersion Relation and Resonance Condition}

\begin{figure}[htb]
\begin{center}
\includegraphics[height=8.6cm]{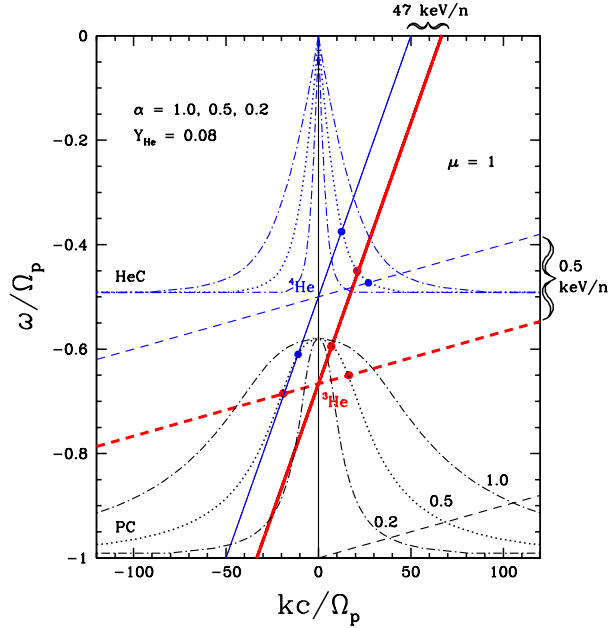}
\end{center}
\caption{ 
Dispersion relations for left-handed polarized parallel propagating waves in the HeC and PC 
branches for ${\rm Y}_{He}=0.08$ and for $\alpha = $1.0 ({\it dot-dashed curves}), 0.5 ({\it 
dotted curves}) and 0.2 ({\it dot-dashed curves}). The solid and dashed lines give the resonance 
conditions for $^3$He ({\it thick lower lines}) and $^4$He ({\it thin upper lines}) with $\mu = 
1.0$ and $E =$ 47 keV/nucleon and 0.5 keV/nucleon. The resonant points of these particles with 
the PC and HeC branches in the plasma with $\alpha = 0.5$ are designated by the filled circles. 
The thin dashed line near the bottom indicate the resonance condition for protons with $\mu = 
1.0$ and $E = 0.5$ keV.
\label{fig1.ps}
}
\end{figure}

To simplify the investigation, we assume that the turbulent plasma waves propagate in a ``cold'' 
fully ionized plasma\footnote{As argued in PL04, the inclusion of thermal effects does not 
change the main conclusions of SA models. These effects in most cases will be overshadowed by the 
uncertainties in the model parameters describing the spectrum of the turbulence.}. Then the 
dispersion relation is determined by the ion abundance and the plasma parameter $\alpha$: the ratio 
of the electron plasma frequency $\omega_{pe}$ to its gyro-frequency $\Omega_{e}$
\begin{equation}
\alpha = \omega_{pe}/\Omega_{e} 
= 3.2 (n_{e}/10^{10}{\rm cm}^{-3})^{1/2}(B_0/100{\rm \ G})^{-1}\,,
\end{equation}
where $n_{e}$ is the electron number density, and $B_0$ is the large scale magnetic field. In the 
case of solar flare plasmas, modifications to the dispersion relation due to $^3$He 
and elements heavier than $^4$He are negligible (Steinacker et al. 1997).  For waves propagating 
parallel to the large scale magnetic fields, one therefore has
\begin{equation}
{k^2\over\omega^2}=1 -{\alpha^2\over \omega\delta}\left[{1\over\omega\delta-1}+{(1-2{\rm Y}_{\rm
He})\over\omega+1}+{{\rm Y}_{\rm He}\over \omega+1/2}\right]\,,
\end{equation}
where the fraction of $^4$He number abundance ${\rm Y}_{\rm He}=0.08$ and $\delta = m_e/m_p$ is the 
electron-to-proton mass ratio. The wave frequency and wavenumber are given by $\omega$ and $k$ 
in units of the proton gyro-frequency $\Omega_{p}$ and $\Omega_p/c$, respectively, where $c$ is 
the speed of light. The left-handed polarized waves are designated with a negative frequency, 
and there are five wave branches (PL04).  Figure \ref{fig1.ps} shows the $^4$He-cyclotron (HeC) 
and proton-cyclotron (PC) branches, which dominate the SA of $^3$He and $^4$He (LPM04), for 
three values of $\alpha$. 

The resonance condition for particles interacting with parallel propagating waves is given by
\begin{equation}
\omega - k\beta \mu = -{\Omega_{i}\over \Omega_p\gamma}\,,
\label{reson}
\end{equation}
where $\beta c$, $\mu$ and $\gamma$ are the particle velocity, cosine of the pitch angle and 
Lorentz factor, respectively, and $\Omega_i$ is the ion gyro-frequency. 
The straight solid and dashed lines in Figure \ref{fig1.ps} indicate this relation for $^3$He 
({\it thick lower lines}) and $^4$He ({\it thin upper lines}) with $\mu = 1.0$ and two energies. The 
intersection points of these 
lines with the dotted lines for the dispersion relation with $\alpha = 0.5$ designate waves that 
satisfy the resonance condition.  The Fokker-Planck coefficients for the particles are 
calculated by adding contributions from these waves (Dung \& Petrosian 1994). The dashed line 
near the bottom gives the resonance condition for 0.5 keV protons with $\mu=1$, which will be useful 
in understanding the thermal damping effects discussed below.

\subsection{Turbulence Spectrum}

There has been limited work on a complete kinetic theory description of plasma turbulence 
detailing the wave generation, cascade, and eventually damping by background particles at small 
scales. Most of the investigations of plasma turbulence are limited to the MHD regime (Bieber et al. 
1994;  Dr\"{o}ge 2003; Hirose et al. 2004), where the wave frequency is much lower than the particle 
gyro-frequency, and the turbulence cascade is not isotropic for most of the wave modes (Cho \& 
Lazarian 2003, 2004). A formulation of these processes for application to solar flares is still 
underdevelopment. Here we study the SA by assuming a power spectrum for the PWT, taking into account 
the knowledge obtained from recent observations of plasma turbulence in the interplanetary medium and 
from theoretical investigations. The main features of PWT can be formated by two length scales: 
the turbulence generation and the wave damping length scale, and the related spectral indexes.

The turbulence generation length scale is usually related to the large scale dynamical
evolution of the system, such as the size of the current sheet between reconnecting magnetic
fields. For high frequency wave branches, such as the PC branch mentioned above, waves can
also be generated via coupling with low frequency MHD waves, like those in the HeC branch (Xie,
Ofman, \& Vi\~{n}as 2004).  
Although the magnetic reconnections and the consequence energy release and turbulence
generation processes seem to involve complex micro- and macro- scopic physics, the turbulence
generation length size must be smaller than the typical dimension of the system. For flare 
conditions, such a scale generally implies low wavenumber or frequency waves, which resonate with 
relativistic particles that is not quite relevant to observations of SEPs with {\it ACE}. This is 
the case for the HeC branch so that the low wavenumber cutoff of this branch simply determines
the energy content in the turbulence. The PC branch, on the other hand, mostly interacts with
nonrelativistic particles. As discussed in PL04 and LPM04, large scale waves in this branch are very
efficient accelerators of hundreds of MeV protons, $^3$He and $^4$He ions. Observations of
SEPs in this energy range can be used to constrain the wave generation at these length scales.  
We therefore take the low wavenumber cutoff $k_{\rm min}$ of the PC branch as one of the primary 
model parameters.

The high wavenumber cutoff of the turbulence spectrum is determined by the damping rate of these 
waves by the background plasma. Thermal damping involves complicated 
nonlinear effects (Swanson 1989, pp. 122, 291, and 305), which can change the wave damping rate, 
modulate the evolution of the wave spectrum at high $k$, and the particle distribution in the 
tail of a thermal background plasma. These determine the injection process of SA. Thermal 
damping of parallel propagating waves by the background ions has been studied in detail by 
Steinacker et al. (1997). One of the major findings of this work is that waves that resonate 
with the high-energy tail of a thermal distribution are subjected to strong Landau damping. 
Because the abundances of $^3$He and elements heavier than $^4$He are very low, their 
contributions to the thermal damping effects can be ignored. But for the short scale waves in 
the HeC and PC branches, which are damped by the background $^4$He and protons, respectively, 
the decay time could be comparable to the proton gyro-period. A steep turbulence spectrum is 
expected in this wavenumber range. 

We assume that the turbulence is unpolarized and has a broken power-law spectrum for each of five 
wave branches:
\begin{equation}
{\cal E}(k) = (q-1){{\cal E}_0\over k_{\rm min}}
\cases
{(k/k_{\rm min})^{q_{\rm l}}, & for $k<k_{\rm min}$, \cr
(k/k_{\rm min})^{-q}, & for $k_{\rm min}<k<k_{\rm max}$, \cr
(k_{\rm min}/k_{\rm max})^q(k/k_{\rm max})^{-q_{\rm h}},  & for $k>k_{\rm max}$,}   
\end{equation}
where ${\cal E}_0$ indicates the intensity of the turbulent plasma waves, and $k_{\rm min}$ 
must be larger (perhaps much larger) than $k_{\rm L}\equiv 2\pi c/L\Omega_p$ for a system with size 
$L$. Note that ${\cal E}_0$, $k_{\rm min}$ and $k_{\rm max}$ could be different for different wave 
branches. However, we assume that the spectral indexes and $\overline{\cal E}_0\equiv (q-1){\cal 
E}_0(k_{\rm min}\delta)^{q-1}$ are the same for all branches so that the wave intensity only depends 
on the wavenumber $k$ in the inertial ranges, i.e., between $k_{\rm min}$ and $k_{\rm max}$. From the 
resonance conditions for $^4$He and protons with $E = 0.5$ keV/nucleon and $\mu = 1.0$ in Figure 
\ref{fig1.ps}, we set the high wavenumber cutoff $k_{\rm max} = 2\alpha\delta^{-1/2}$ for the PC 
branch and one half of this value for the HeC branch. This characterizes the damping of waves under 
plasma conditions typical for solar flares (see Xie et al. 2004). Because the PC branch is the 
dominant branch here, in what follows the corresponding quantities refer to this branch unless 
specified otherwise. We choose $q = 2.0$ and $q_{\rm h}=4.0$ as the fiducial parameters, which are 
consistent with solar neighborhood observations (Bieber et al. 1994; Dr\"{o}ge 2003)\footnote{The 
value of $q_{\rm l}$ is unimportant as long as it is much larger than 1. We set $q_{\rm l}=2$ in our 
calculation.}. For $q=2.0$ the acceleration and scattering rates of relativistic particles are 
independent of energy (corresponding to the so-called hard sphere approximation), which gives rise to 
a power-law particle distribution that is cut off at an energy where the particle resonates with 
waves with $k\le k_{\rm min}$, or where a loss process becomes dominant (PL04). Clearly the total 
turbulence energy density ${\cal E}_{\rm tot}\simeq 2{\displaystyle \sum_{\sigma}} {\cal 
E}^\sigma_0=[2\overline{\cal E}_0/(q-1)]{\displaystyle \sum_{\sigma}}(k_{\rm 
min}^{\sigma}\delta)^{1-q)}$, where 
$\sigma$ indicates the wave branches and the factor of 2 arises from the two propagation directions 
of the waves. 

\subsection{Fokker-Planck Coefficients and Scattering and Acceleration Times}

With the turbulence spectrum specified, one can proceed to calculate the Fokker-Planck 
coefficients (Dung \& Petrosian 1994):
\begin{equation}
\left.
\begin{array}{l} 
D^i_{\mu\mu} \\
D^i_{\mu p} \\
D^i_{pp} 
\end{array}
\right\} 
={\Omega_i^2\delta(\mu^{-2}-1)
\over 2 \Omega_p^2 \gamma^2\tau_p}\sum_{j=1}^N\chi(k_j)
\cases{\mu\mu (1-x_j)^2,  \cr
\mu p x_j(1-x_j), \cr
p^2 x_j^2, \cr}
\label{coeff}
\end{equation}
where
\begin{equation}
\chi(k_j) = {{\cal E}(k_j)/\overline{\cal E}_0\over|\beta 
\mu-\beta_g(k_j)|}\,,\,\,\,\,\,\,\,\,\,
\,\,\,\,\,\,\,\, x_j=\mu\omega_j/\beta k_j\,,
\label{chi}
\end{equation}  
$p$ and $\Omega_i$ are the particle momentum and gyro-frequency, 
respectively, and $\beta_g = {\rm d}\omega/{\rm d}k$ is the group velocity 
of the wave. The sum over $j$ is for the resonant points discussed above. Following previous 
studies (Dung \& Petrosian 1994; Pryadko \& Petrosian 1997; PL04), we describe the turbulence 
intensity with a characteristic timescale $\tau_p$
\begin{equation}
\tau_p^{-1} = \pi\Omega_e\left[{\overline{\cal E}_0\over
B_0^2/8\pi}\right]\,\, \ \ \ \ {\rm with} \ \ \ \ \overline{\cal E}_0 = {(q-1){\cal E}_{\rm tot}\over 
2{\displaystyle \sum_\sigma}(k^\sigma_{\rm min}\delta)^{1-q}}.
\label{taup}
\end{equation}  

\begin{figure}[htb]
\begin{center}
\includegraphics[height=8.4cm]{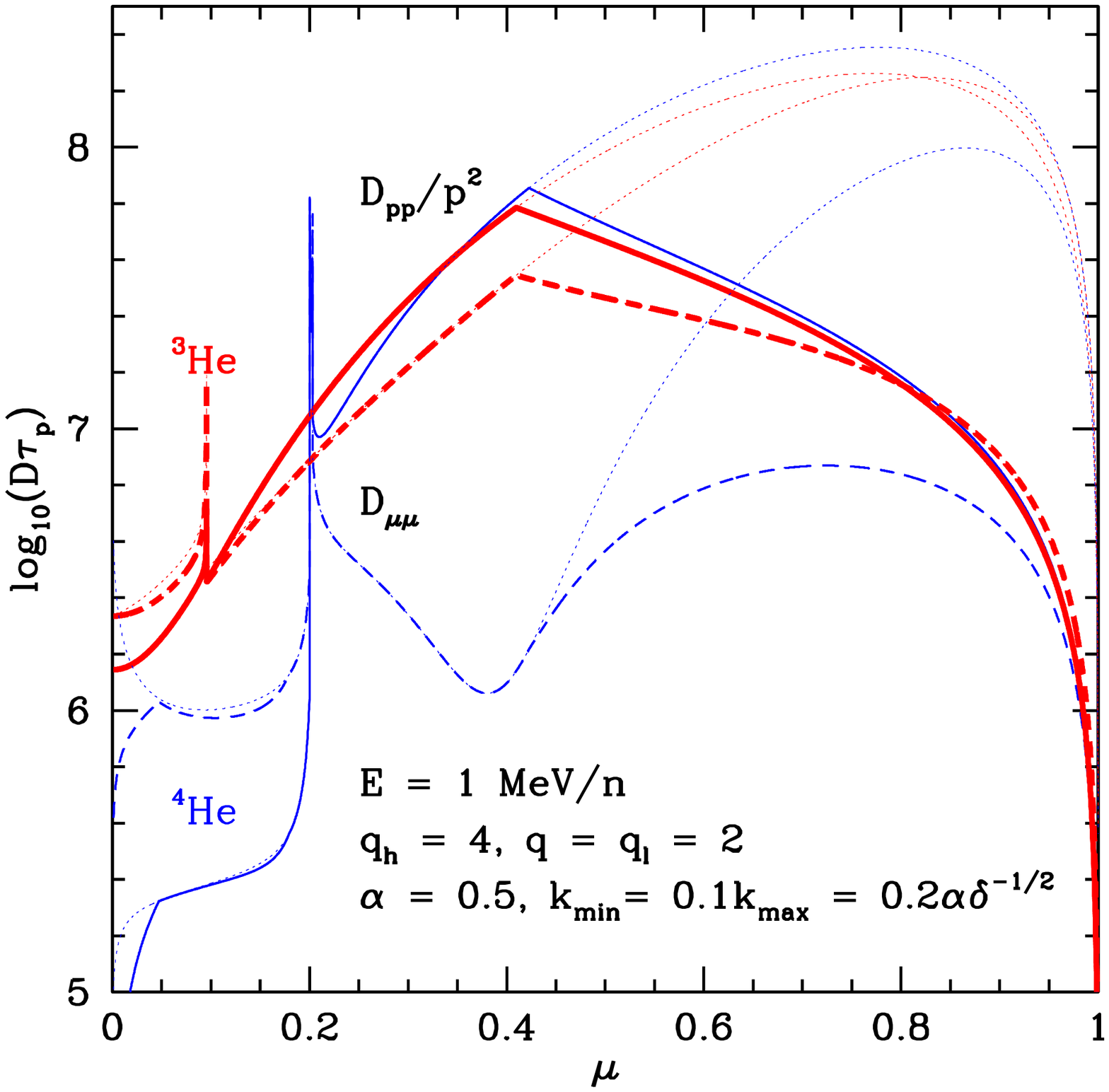}
\hspace{-0.6cm}
\includegraphics[height=8.4cm]{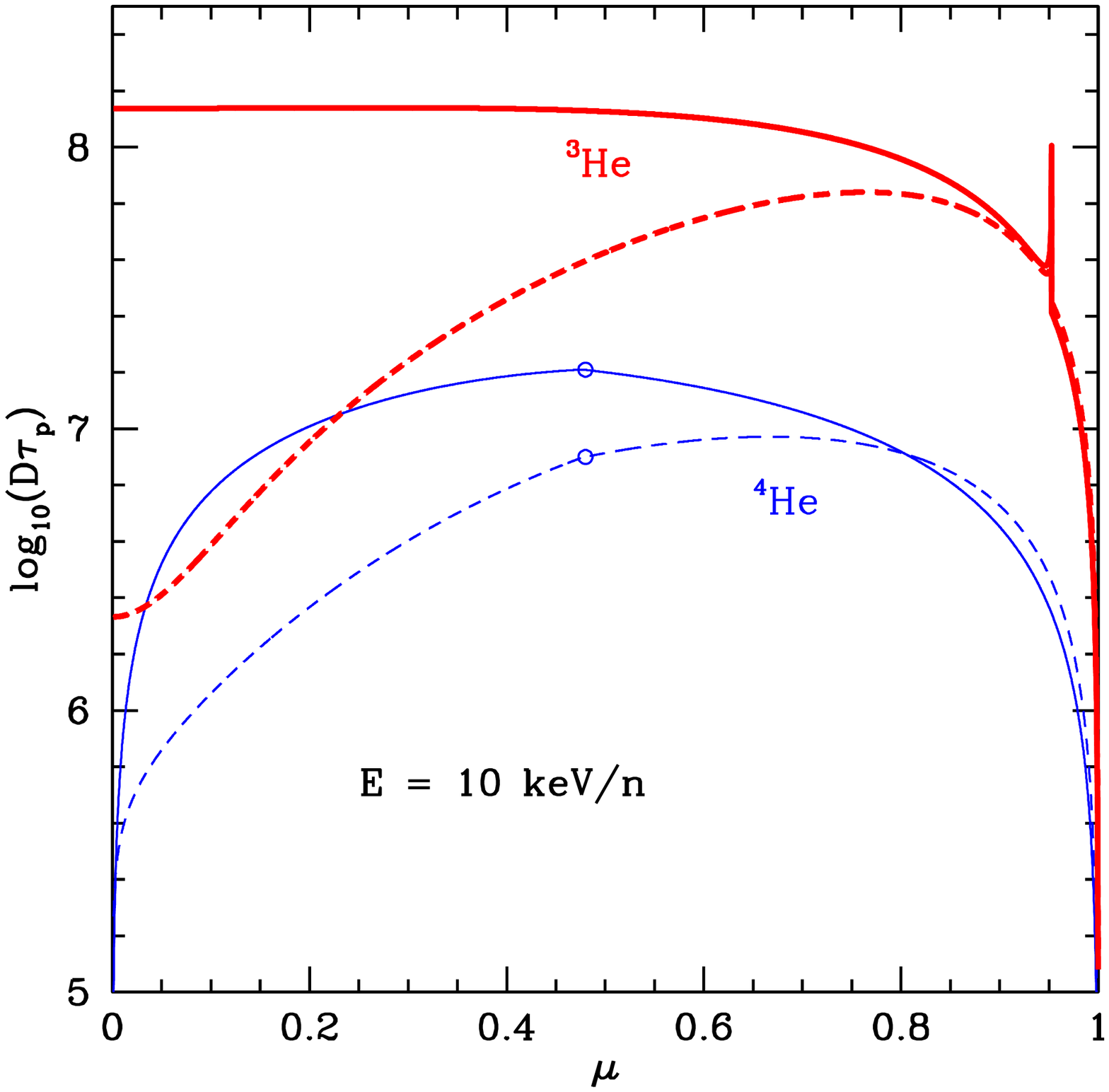}
\end{center}
\caption{
{\it Left:} Pitch angle dependence of the dimensionless scattering ($D_{\mu\mu}\tau_p$: {\it 
dashed lines}) and acceleration ($\tau_p D_{pp}/p^2$: {\it solid lines}) rates of 1 MeV/nucleon 
$^3$He ({\it thick lines}) and $^4$He ({\it thin lines}) in a plasma with $\alpha = 0.5$. 
Model parameters are indicated in the figure. The thin dotted lines give the corresponding rates for 
a pure power-law turbulence spectrum with $k_{\rm max}\to \infty$ and $k_{\rm min}\to k_{\rm L}$, or 
$q = q_{\rm h} = 2$ for the model. The small differences at small $\mu$ from
the model with a general broken power-law spectrum are due to the thermal damping effects. The 
large differences at large $\mu$ are due to the low wavenumber cutoff $k_{\rm min}$. One sees 
that the introduction of $k_{\rm min}\ll k_{\rm L}$ can reduce the interaction rates of particles at 
this energy significantly and the acceleration rate is usually higher than the corresponding 
scattering rate. {\it Right:} Same as the left panel, but for $E = 10$ keV/nucleon particles. The 
results for a pure power-law turbulence spectrum are not shown. Here the acceleration rate of $^3$He 
is more than one order of magnitude higher than that for $^4$He. The open circles indicate breaks 
caused by $k_{\rm max}$.
}
\label{fig2.ps}
\end{figure}

Figure \ref{fig2.ps} gives the dimensionless acceleration rate $(D_{pp}/p^2)\tau_p$ ({\it solid 
lines}) and scattering rate $D_{\mu\mu}\tau_p$ ({\it dashed lines}), as functions of $\mu$, for 
$^3$He ({\it thick lines}) and $^4$He ({\it thin lines}) at $E =$ 1 MeV/nucleon ({\it left} 
panel) and 10 keV/nucleon ({\it right} panel) in a plasma with $\alpha = 0.5$. The low 
wavenumber cutoff of the PC branch $k_{\rm min} = 0.1 k_{\rm max} = 0.2 \alpha \delta^{-1/2}$ 
here. The values of the other model parameters introduced above are indicated in the figure. It 
is obvious that the scattering rate is usually lower than the corresponding acceleration rate. 
Because of this, to the extent that the acceleration rate depends on $\mu$, the particle 
distribution will be anisotropic, unless there are other scattering agents. More importantly, the 
acceleration rates of $^3$He and $^4$He are comparable at 1 MeV/nucleon, but the $^4$He 
acceleration rate is more than 10 times lower than that of $^3$He at lower energies, e.g. at 10 
keV/nucleon (see the {\it right} panel). This is because $D_{pp}$ is proportional to the phase 
velocity square of the resonant waves and at low energies waves resonating with $^4$He have 
relatively lower phase velocities.  Thus, compared with $^3$He, the acceleration of $^4$He from 
a low energy can be suppressed significantly.

\clearpage

To illustrate the effects of introducing the high and low wavenumber cutoffs, the left panel 
also shows the rates for a pure power-law turbulence spectrum ($k_{\rm max}\to \infty$ and 
$k_{\rm min}\to k_{\rm L}$), i.e. $q = q_{\rm h} = 2$, as indicated by the thin dotted lines. 
From the resonance condition (\ref{reson}) one can show that with the thermal damping and 
reduction of the intensity of high $k$ waves the interaction rates at small $\mu$ decreases. 
The more pronounced differences at large $\mu$, on the other hand, are mostly due to the low 
wavenumber cutoff $k_{\rm 
min}$. One should also notice that the thermal damping effects reduce the $^4$He rates much more 
than that of $^3$He. This can be clearly seen in the right panel, where $k_{\rm 
max}$ introduces a break in the $^4$He interaction rates near $\mu = 0.5$. No such features are 
seen in the rates for $^3$He. This is expected because $^3$He mostly interacts with large scale 
waves that are not damped. The sharp spikes in the interaction rates are due to the critical 
angles studied in PL04, when the denominator of equation (\ref{chi}) approaches zero (for
interactions with the PC branch). These spikes are very narrow and therefore do not contribute
significantly to the pitch angle averaged interaction rates. However, once crossing these
spikes, particles start to interact with waves with different phase velocities, and the
interaction rate can change dramatically. The left panel illustrates this clearly. At small
$\mu$ the acceleration rate of $^4$He is about ten times lower than that of $^3$He.  Beyond
the spike at $\mu\sim 0.2$ (for $^4$He), the two acceleration rates become comparable since the
particles interact with waves with similar frequencies and wavenumbers.

\begin{figure}[htb]
\begin{center}
\includegraphics[height=8.6cm]{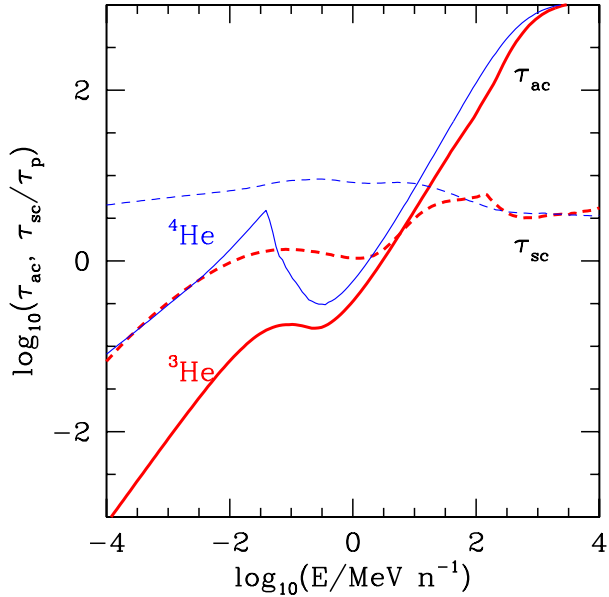}
\end{center}
\caption{
The energy dependence of pitch angle averaged scattering ({\it dashed lines}) and 
acceleration ({\it solid lines}) times for $^4$He ({\it thin lines}) and $^3$He ({\it thick 
lines}). The model parameters are the same as those in Figure \ref{fig2.ps}.
\label{fig3.ps}
}
\end{figure}

Because the acceleration rate is usually higher than the scattering rate at low energies, one 
can define the pitch angle averaged acceleration and scattering times (Pryadko \& Petrosian 
1997):
\begin{eqnarray}
\tau_{\rm ac} &=& {2p^2\over
\int_{-1}^1{\rm d}\mu D_{pp}(\mu)}\,, \label{ac} \\ 
\tau_{\rm sc} &=& \int_{-1}^1{\rm d}\mu(1-\mu^2)^2/D_{\mu\mu}\,. \label{sc}
\end{eqnarray}
Note that here we define $\tau_{\rm ac}$ differently from that in LPM04 where we assumed isotropy and 
that $D_{\mu\mu}\gg D_{pp}/p^2$, which is always the case at high energies. But as shown above this 
is not true at low energies so that the above relation is more accurate. This makes only a 
quantitative (not a qualitative) difference. Figure \ref{fig3.ps} shows the energy dependence of 
these times (acceleration time: {\it solid lines}, scattering time: {\it dashed lines}) for $^3$He 
({\it thick lines}) and $^4$He ({\it thin lines}) for the model depicted in Figure \ref{fig2.ps}. The 
scattering time is longer than the corresponding acceleration time below a few MeV/nucleon. First, 
this implies that at low energies SA is more efficient than diffusive shock acceleration, whose 
acceleration time is comparable to the scattering time. We also expect some anisotropy in accelerated 
particle distributions with a pitch angle distribution similar to the pitch angle dependence 
of the acceleration rates. However, as shown in Figure \ref{fig2.ps} the expected anisotropy will not 
be large, and other scattering processes, such as Coulomb collisions, may further isotropize the 
accelerated particle distribution. Beyond a few tens of MeV/nucleon scattering becomes more 
efficient, and isotropy of the particle distribution is established by wave scatterings. 

Below a few tens of keV/nucleon the acceleration time of $^4$He is more than one order of 
magnitude longer than that of $^3$He. The sharp decrease of the $^4$He acceleration time 
near 40 keV/nucleon is due to the interaction of $^4$He with large scale waves in the PC 
branch, which also dominate the acceleration of $^3$He (see Fig. \ref{fig1.ps}). The 
acceleration times of the two particles become comparable beyond 100 keV/nucleon and are 
identical at relativistic energies. 

\subsection{Dependence of Acceleration Times on Model Parameters}

The physical conditions for different solar flares are expected to be different.  In this 
section we investigate how the acceleration times depend on the model parameters.

\begin{figure}[htb]
\begin{center}
\includegraphics[height=8.4cm]{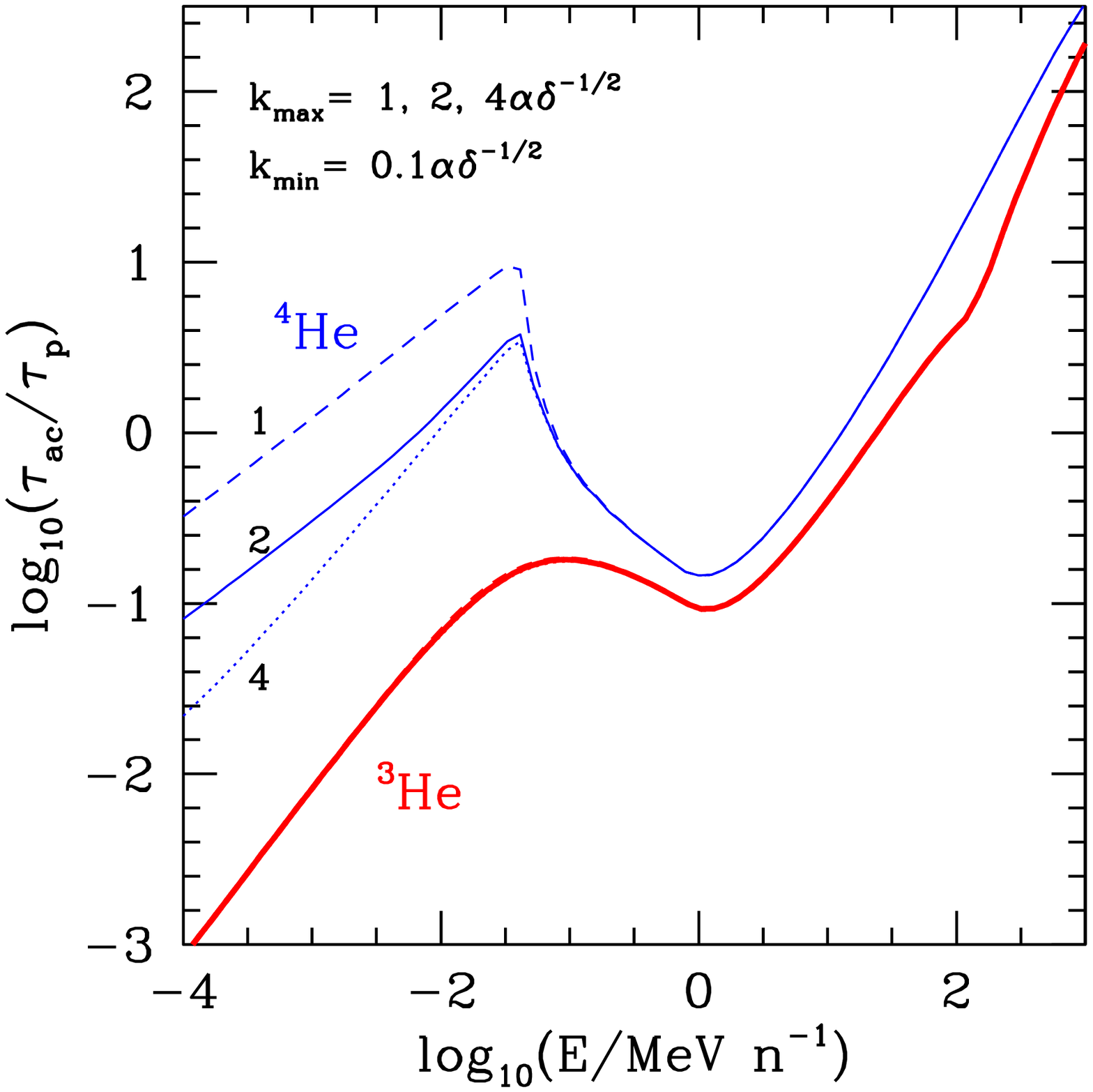}
\hspace{-0.6cm}
\includegraphics[height=8.4cm]{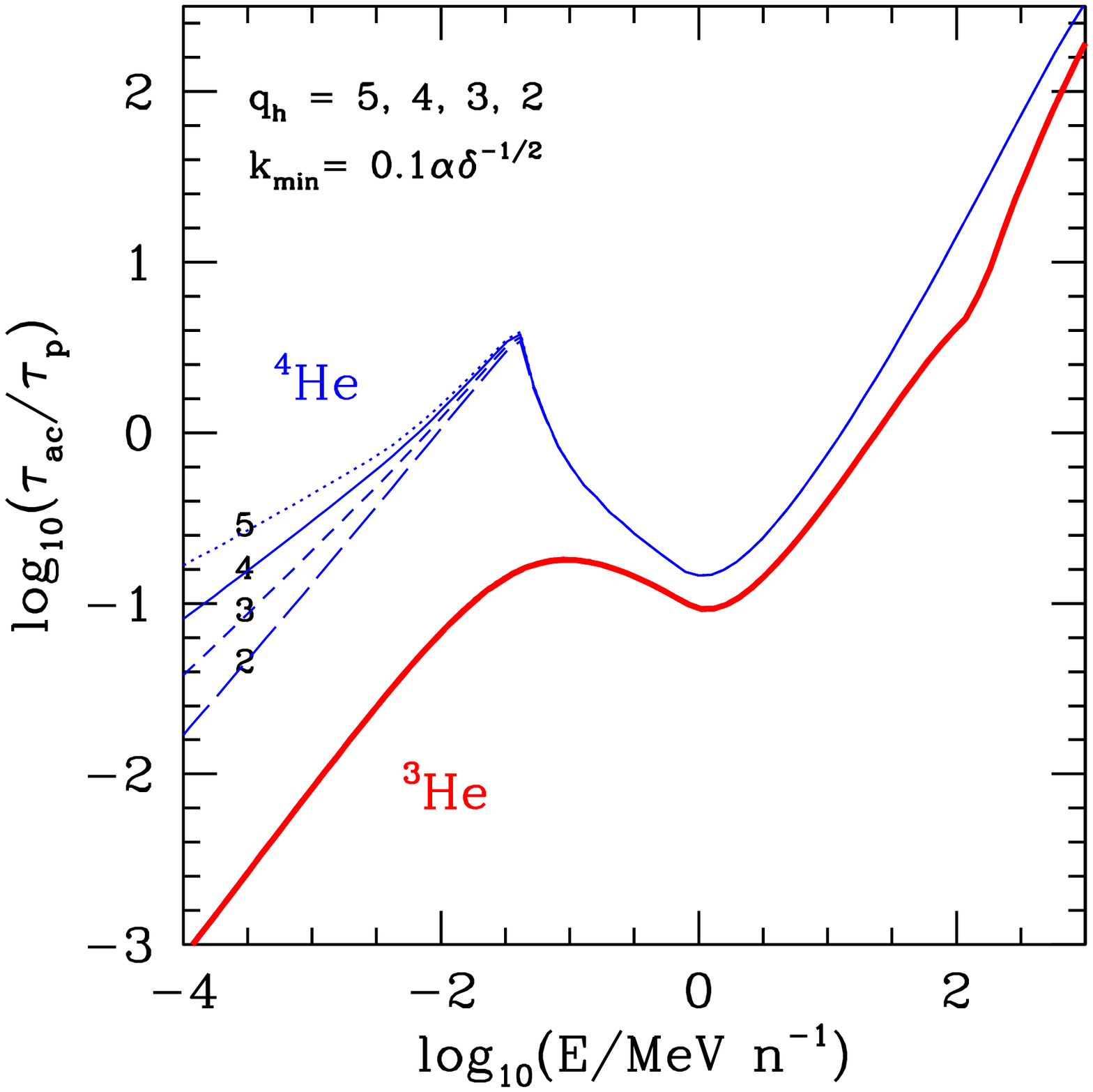}
\end{center}
\caption{
{\it Left:} Dependence of the acceleration time of $^4$He on the high wavenumber cutoff $k_{\rm 
max}$ with $k_{\rm min} = 0.1\alpha\delta^{-1/2}$. Note that acceleration of $^3$He is not affected. 
Other model parameters are the same as those in Figure \ref{fig2.ps}.  The solid lines are similar to 
those in Figure \ref{fig3.ps} except for the difference due to a lower $k_{\rm min}$ adopted here. In 
what follows we choose this as the fiducial model (always indicated by solid lines). The high 
wavenumber cutoff for the PC branch is always two times higher 
than that for the HeC branch. The lines are labeled with their corresponding $k_{\rm max}$ in 
 units of $\alpha \delta^{-1/2}$. The dashed line is for $k_{\rm max} = \alpha \delta^{-1/2}$ and the 
dotted lines for a four times 
higher $k_{\rm max}$. 
{\it Right:} Same as the left panel except that here 
the dependence on the spectral index $q_{\rm h}$ is shown. Note that the $q_{\rm h}=2$ model has no 
thermal damping.
}
\label{fig4.ps}
\end{figure}

As discussed above, we parameterize the wave spectrum by several parameters with two of them 
characterizing the thermal damping; namely the high wavenumber cutoff $k_{\rm max}$ and the 
turbulence spectral index $q_{\rm h}$. Figure \ref{fig4.ps} shows how the acceleration 
times of $^3$He and $^4$He change with these parameters. The left panel is for different 
$k_{\rm max}$, and the right panel has different $q_{\rm h}$. Here the solid lines represent our 
fiducial model, which is analogous to that in Figure \ref{fig3.ps} except that the low 
wavenumber cutoff $k_{\rm min}$ is now two times lower: $k_{\rm min} = 0.1\alpha\delta^{-1/2}$. 

\begin{figure}[htb]
\begin{center}
\includegraphics[height=8.4cm]{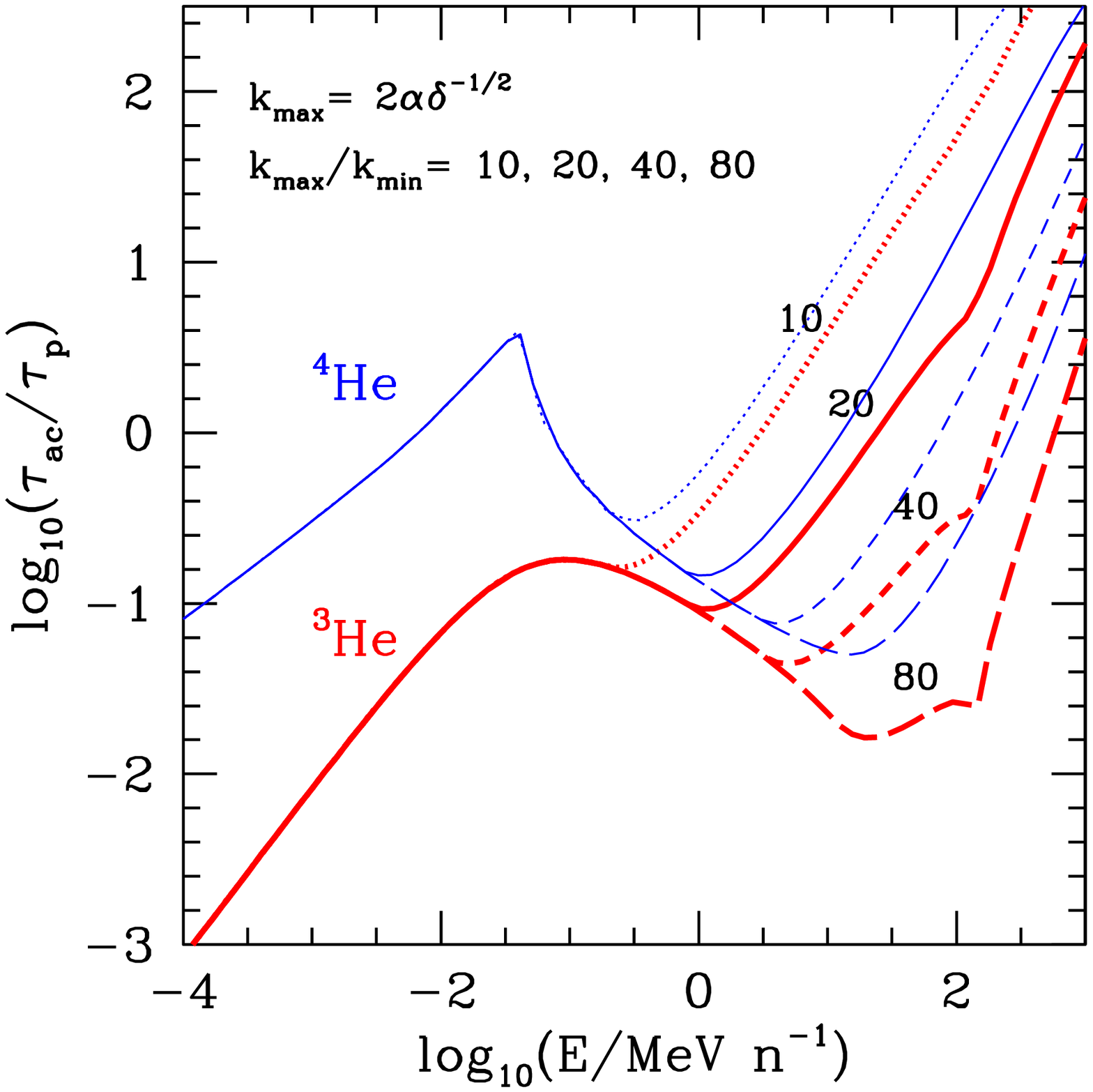}
\hspace{-0.6cm}
\includegraphics[height=8.4cm]{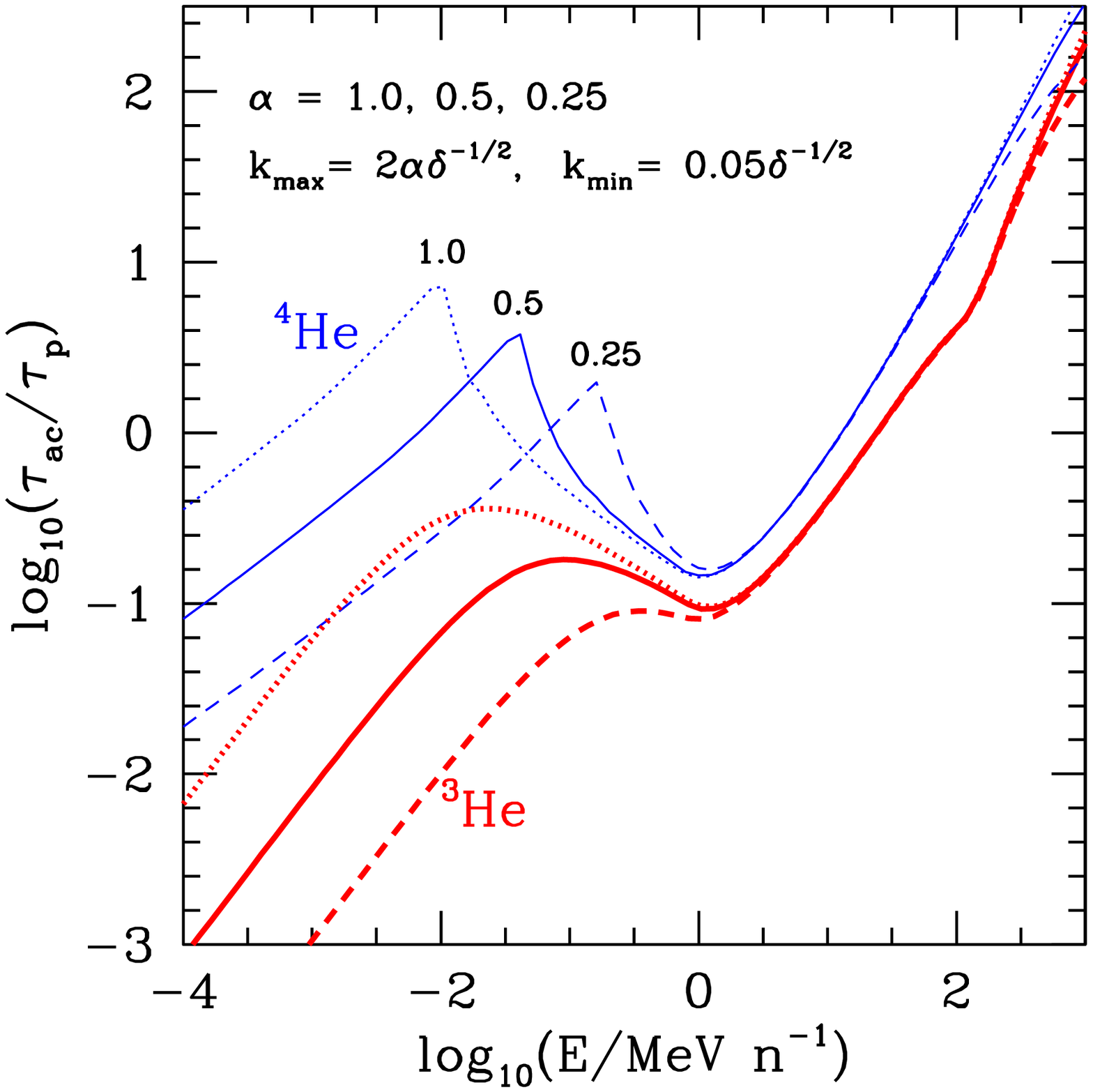}
\end{center}
\caption{
Same as Fig. \ref{fig4.ps} but showing the dependence of the acceleration times on $k_{\rm 
min}$ ({\it left}) and the plasma parameter $\alpha$ ({\it right}). Here $k_{\rm max}$ changes 
with $\alpha$ (See Fig. \ref{fig1.ps}), but all other model parameters remain the same as in 
the fiducial model. 
}
\label{fig5.ps}
\end{figure}

One obvious feature is that the thermal damping does not affect the acceleration time of $^3$He.
As mentioned above, this is because $^3$He mostly interacts with large scale waves in the PC
branch, which can not be damped by the low-energy protons and $\alpha$-particles in the
background. The lower $k_{\rm max}$ and the larger $q_{\rm h}$, the stronger the thermal 
damping, and, as expected, the longer the acceleration time of $^4$He. Higher energy particle 
acceleration time is not affected by the damping because these particles interact with
low wavenumber waves. We also note that the acceleration time of $^4$He at low energies is more
sensitive to $k_{\rm max}$ than to $q_{\rm h}$. Even in the case $q_{\rm h} = 2$, which
corresponds to the absence of damping effects, the $^4$He acceleration time is still more than
ten times longer than the $^3$He acceleration time at low energies. So the $^4$He acceleration
from a low energy can be suppressed even without the thermal damping effects. This suppression
is purely because of the modification to the dispersion by the presence in the plasma of a 
significant population of $\alpha$-particles, and only depends on the plasma parameter $\alpha$ 
and the $^4$He abundance ${\rm Y}_{\rm He}$.

\begin{figure}[htb]
\begin{center}
\includegraphics[height=8.6cm]{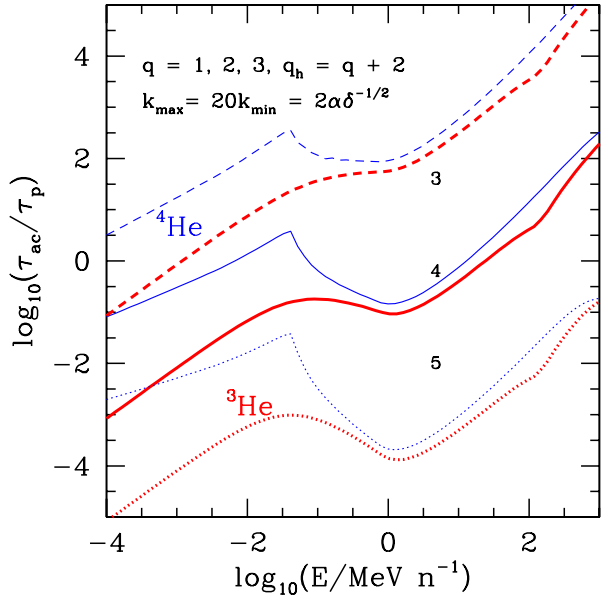}
\end{center}
\caption{
Same as Fig. \ref{fig4.ps} except for the dependence of the acceleration times on the turbulence 
spectral indexes $q$ and $q_{\rm h}$. $q_{\rm h} - q = 2$. All other model parameters remain 
the same as the fiducial model. 
}
\label{fig6.ps}
\end{figure}

As mentioned earlier, the generation length of waves in the PC branch, corresponding to $k_{\rm
min}$, is another important parameter. The left panel of Figure \ref{fig5.ps} shows how $k_{\rm
min}$ affects the acceleration times. (The low wavenumber cutoffs of other branches are assumed
to be far lower so that they do not affect the acceleration of nonrelativistic particles.)
Obviously these large scale waves mostly interact with particles in the MeV/nucleon energy
range, so that the acceleration times at low energies remain unchanged. However the acceleration
of MeV/nucleon particles is very sensitive to this parameter for both $^3$He and $^4$He. With the 
decrease of $k_{\rm min}$ particle acceleration at high energies becomes more efficient but the 
relative rate of the two ions changes very little. In the following sections we show
that the high-energy cutoffs of the accelerated particle spectra are very sensitive to this
parameter and observations with {\it ACE} can be used to constrain it directly.

The right panel of Figure \ref{fig5.ps} shows the dependence of the acceleration time on the
plasma parameter $\alpha$. The most prominent feature of these curves is the decrease of $\tau_{\rm 
ac}$ for both ions at low energies with the decrease of $\alpha$. This does not affect the $^3$He 
acceleration but has important effects on the $^4$He acceleration because of its long $\tau_{\rm 
ac}$ (see \S\ \ref{dis}). Because $k_{\rm min}$ is fixed, the acceleration times in the MeV/nucleon
energy range do not change significantly. The plasma parameter, therefore, mostly affects the
acceleration rates at low energies and the relative acceleration of the ions. 

Figure \ref{fig6.ps} shows how the acceleration times change with the turbulence spectral 
indexes $q$.  Here we keep $q_{\rm h}-q = 2$ and other model parameters unchanged in these 
calculations. As expected, the acceleration of high-energy 
particles becomes more efficient when the turbulence spectrum becomes steeper since high-energy 
particles mostly resonate with large scale waves (eq. [\ref{reson}]). (Note that the 
normalization time $\tau_p$ changes with $q$.) Surprisingly, however, for 
all the model parameters considered here the acceleration time of $^4$He is always longer 
than the $^3$He acceleration time. This is one of the key features of SA by parallel 
propagating waves and has profound implications on the enhancement of $^3$He and the relative 
acceleration of the two species. 

\section{MODEL DESCRIPTION AND APPLICATION TO SEPS}
\label{sep}

In the SA theory ions gain energy via their resonant interactions with the PWT and lose 
energy via Coulomb collisions with the background thermal particles. These processes can be 
described by the Fokker-Planck equation, whose coefficients have been calculated in the previous 
section for interactions with parallel propagating waves under the quasi-linear approximation 
and where the perpendicular diffusion across magnetic field lines is ignored. When the wave-ion 
scattering rate is comparable to the ion acceleration rate, in principle one needs to solve the 
four-dimensional equation for the particle phase space distribution with the time $t$, momentum 
$p$, pitch angle cosine $\mu$, and the location along the large scale magnetic field line $s$ as 
variables. 
But the problem can be simplified considerably when one of the rates is much higher than the 
other (Pryadko \& Petrosian 1997). This is the case in the low- and high- energy limits 
considered here, where the acceleration rate and scattering rate dominate, respectively. 
Moreover, if the pitch angle dependence of the diffusion coefficients is weak (e.g. peaked in a 
narrow range of $\mu$, see Fig. \ref{fig2.ps}), then the pitch angle distribution of the 
particles will be relatively smooth and the anisotropy will be small so that one can use pitch angle 
averaged quantities. Furthermore, because we are mostly interested in the spatially integrated 
spectra, we can also integrate the particle distribution spatially (along the field lines). Then the 
acceleration of particles from a low-energy thermal background to relativistic energies can be 
treated fairly accurately by the well-known equation 
\begin{equation}
{\partial N\over\partial t}= {\partial^2\over \partial \E^2}(D_{\E\E} N) + 
{\partial\over \partial \E}[({\dot \E}_L-A(\E)) N] -{N\over T_{\rm esc}} + \dot{Q}\,, 
\label{dceq}   
\end{equation}
where $N(\E)$ is the pitch angle averaged and spatially integrated distribution of accelerated 
particles with kinetic energy $\E$, $D_{\E\E}=\E^2\tau_{\rm ac}^{-1}$ is the 
pitch angle and spatially averaged diffusion rate, and $A(\E) = {\rm d}D_{\E\E}/{\rm d}\E + 
D_{\E\E}(2-\gamma^{-2})/\E(1+\gamma^{-1})$ is the direct acceleration rate. The loss rate 
${\dot \E}_L$ due to Coulomb collisions is summarized in PL04. The escape time $T_{\rm esc} 
= L/2^{1/2}v + L^2/v^2\tau_{\rm sc}$ describes the spatial diffusion of the particles along 
the large scale magnetic field lines. $\dot{Q}$ is a spatially integrated source term, which we 
identify as the thermal background plasma. For quantitative modeling, one needs to specify the length 
of the acceleration region $L$ and the background plasma density $n_e$, magnetic field $B_0$, and 
temperature $T$, which we assume is the same for all background charged particles.

Each term in equation (\ref{dceq}) can be characterized by an energy dependent timescale. The 
timescale of source particle injection $\dot{Q}$ is determined by the large scale dynamical 
evolution of the system, which is comparable to the modulation timescale of the solar flare. 
Because the other interaction times are usually much shorter than the dynamical time of the system, 
one only needs to consider the steady-state solution of the equation with $\dot{Q}=$ constant. Then 
the escaping flux $f = N/T_{\rm esc}$ can be compared with the observed fluences of SEPs
with a normalization factor taking into account the duration of the flare and the cross
sectional area of the flaring magnetic fields. (Note that for the impulsive SEPs studied here,
the transport effects of the ions from the flaring site in the solar corona to {\it ACE} in the
earth's neighborhood are negligible because their scattering mean free path is long [Mason et al.
1989, 2002].) To appreciate spectral features of the corresponding solutions, besides the escape
time mentioned above, we also define a loss time $\tau_{\rm loss} = \E/{\dot \E}_{L}$ and a
direct acceleration time $\tau_{\rm a} = \E/A\sim\tau_{\rm ac}$ for the second term on the
right-hand side of equation (\ref{dceq}). Then the ion loss time as a function of the ion
kinetic energy per nucleon is proportional to the ion atomic number and is inversely
proportional to the square of the ion charge (PL04).

\subsection{Spectral Fit}

\begin{figure}[htb]
\begin{center}
\includegraphics[height=8.4cm]{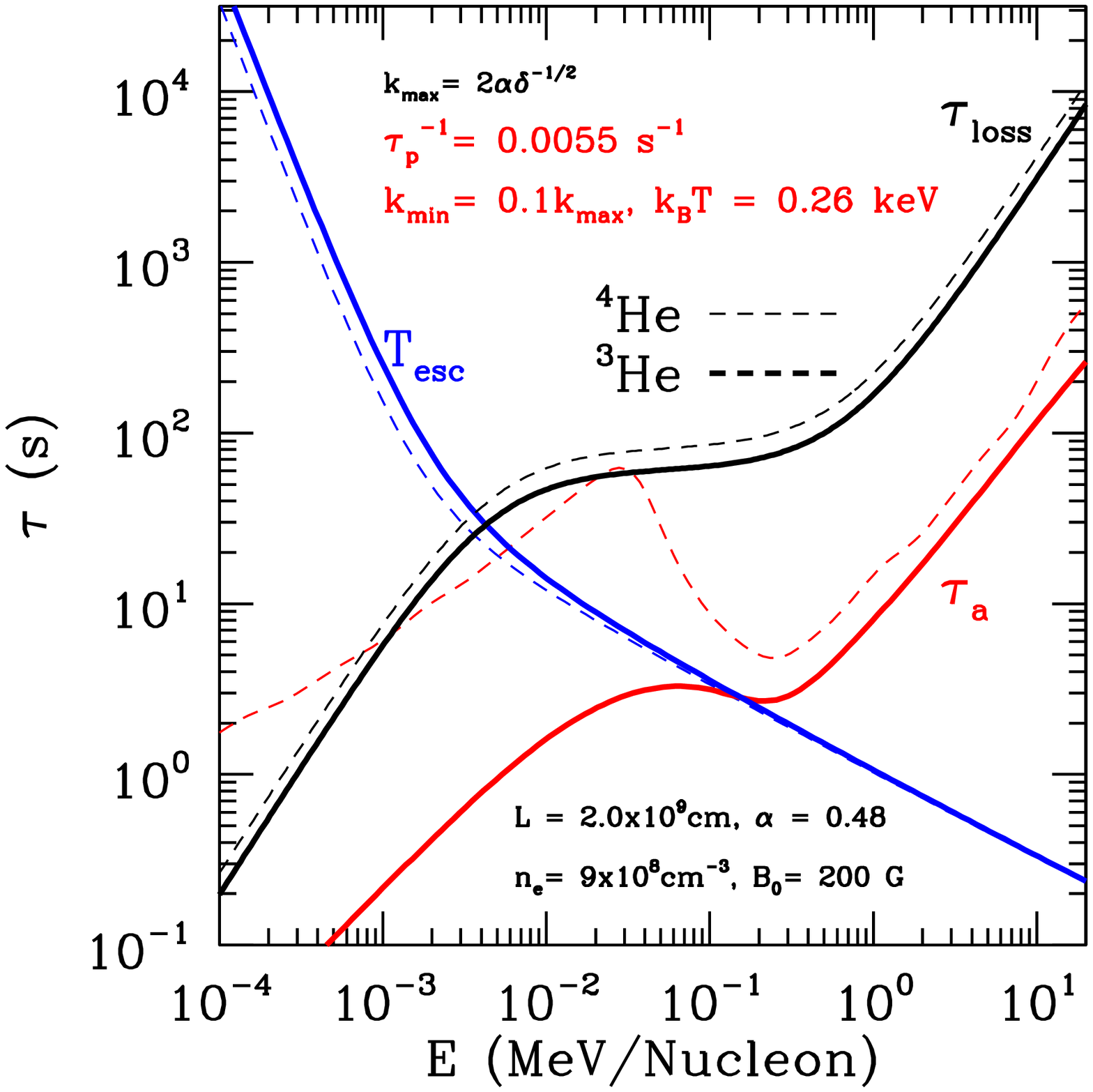}
\hspace{-0.6cm}
\includegraphics[height=8.4cm]{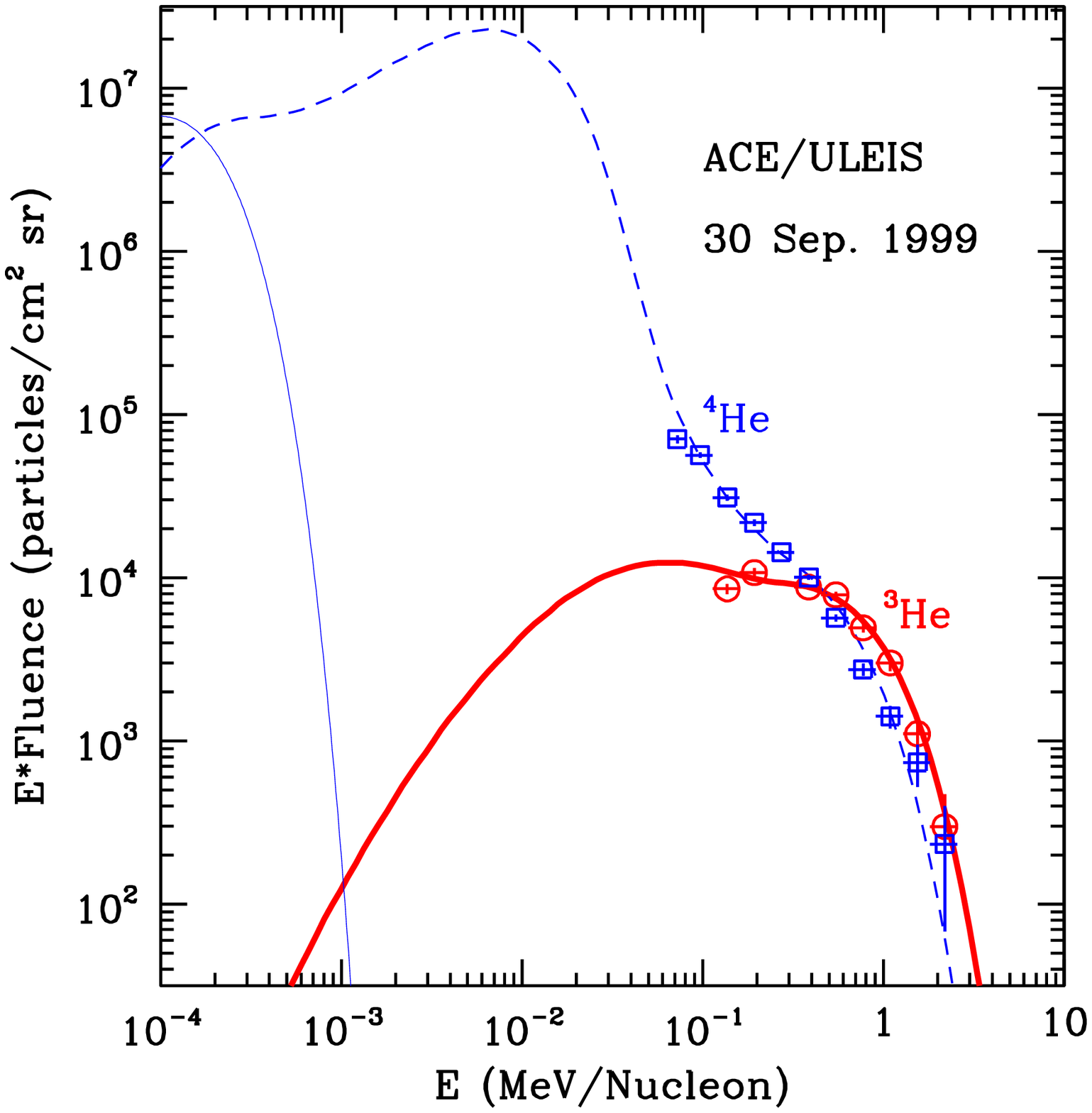}
\end{center}
\caption{
{\it Left:} Direct acceleration ($\tau_{\rm a}$), loss ($\tau_{\rm loss}$), and escape ($T_{\rm 
esc}$) times, as functions of the particle kinetic energy per nucleon, for $^3$He ({\it solid 
lines}) and $^4$He ({\it dashed lines}) ions in a SA model for the event on 1999 September 
30. Model parameters are indicated in the figure. The acceleration times are very similar to 
those of the model in Figure \ref{fig3.ps}. The loss term is dominated by Coulomb collisions 
with background protons below $\sim 10$ keV/nucleon and with electrons above it. The loss time 
in the tens of keV/nucleon energy range is very sensitive to the temperature of the background 
electrons ($\propto T^{3/2}$). The sharp increase of the escape times with the decrease of 
energy below a few keV/nucleon is due to the suppression of the particle spatial diffusion 
rate by efficient Coulomb scatterings, whose rate is comparable to the ion energy loss rate.
Scatterings are unimportant at high energies so that $T_{\rm esc}\simeq L/2^{1/2}v$.
{\it Right:} The corresponding model fit to the observed $^3$He ({\it circiles}) and $^4$He 
({\it squares}) spectra. The thin solid line gives the distribution of the injected thermal 
$\alpha$-particles with arbitrary normalization. (Note that the temperatures of both background ions 
are equal.) 
}
\label{fig7.ps}
\end{figure}

The left panel of Figure \ref{fig7.ps} shows these timescales for a model with $L = 2\times 10^9$ cm, 
$B_0=200$ G, $n_{e}=9\times 10^8$ cm$^{-3}$, the temperature of the background plasma $k_{\rm B} 
T=0.26$ keV, and the characteristic timescale $\tau_{p}^{-1} = 5.5\times 10^{-3}$ s$^{-1}$.  All 
other model parameters are the same as in Figure \ref{fig3.ps}. The corresponding plasma parameter 
$\alpha = 0.48$, and $8\pi \overline{\cal E}_0/B_0^2 = 4.9\times 10^{-13}$. The total turbulence 
energy density can be very low if the generation length scale of 
waves in the HeC branch is much shorter than $L$.  All these parameters are typical for solar flares, 
and a weak level of turbulence also justifies the quasi-linear approximation adopted here and is 
consistent with the general view that the flares are powered by energies stored in the pre-flare 
magnetic fields presumably via magnetic reconnections. The right panel shows the model fit to the 
observed $^3$He and $^4$He spectra of an impulsive SEP on 1999 September 30 observed with {\it ACE}. 
The injected (background) plasma is assumed to have a solar abundance so that the number of $^3$He 
ions integrated over the whole energy range is 2000 times fewer than that of $^4$He.  The model 
clearly produces the observed extreme enhancement of $^3$He and gives a reasonable fit to the spectra 
of both ions.

\clearpage

Because the acceleration time of $^3$He is always much shorter than its Coulomb loss time,
the accelerated spectrum is determined by the acceleration and escape terms.  
Injected at a very low energy, almost all thermal ($k_{\rm B}T=0.26$ keV) $^3$He particles are 
accelerated to the hundreds of keV/nucleon energy range, where the escape process
starts to dominate over the acceleration process. The sharp spectral cutoff near 1
MeV/nucleon results from the quick divergence of the acceleration and the escape times above
this energy.  The SA acceleration is effectively quenched at higher energies. The
acceleration of $^4$He, on the other hand, is quite different. The acceleration time of
$^4$He is comparable with its loss time below a few tens of
keV/nucleon. Near the energy of the injected thermal particles, $A(\E)\le\dot{\E}_{\rm L}$ and
the direct acceleration is difficult, the energy diffusion term on the right-hand side of
equation (\ref{dceq}) dominates the acceleration processes so that more than half of the
injected particles are accelerated to energies $>1$ keV/nucleon. However, the escape time
becomes much shorter than the other times between 10 and 100 keV/nucleon, resulting in a very
steep spectrum beyond $\sim10$ keV/nucleon. Consequently most of the accelerated
$\alpha$-particles stay below $\sim10$ keV/nucleon. This suppresses the
acceleration of $^4$He ions to higher energies and gives rise to the observed enhancement
of $^3$He ions in the MeV/nucleon energy range. At a few hundreds of keV/nucleon the $^4$He
acceleration time becomes comparable with its escape time due to very efficient
acceleration by large scale waves in the PC branch. This gives rise to a relative hard accelerated 
$^4$He spectrum. However, because the $^4$He acceleration time is always longer than that of
$^3$He, its spectrum is always softer than the spectrum of accelerated $^3$He ions.

\begin{figure}[htb]
\begin{center}
\includegraphics[height=5.4cm]{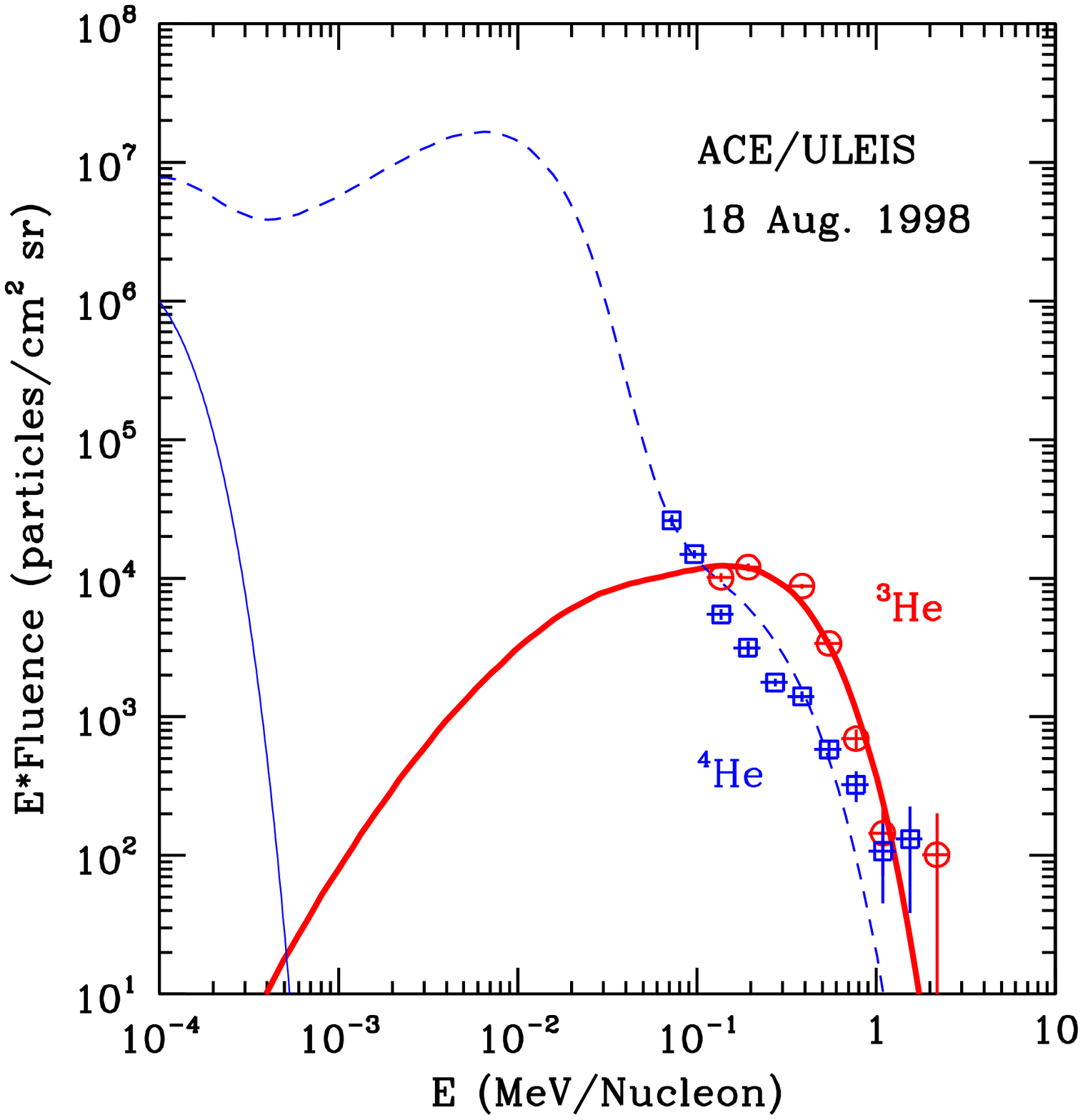}
\hspace{-0.6cm}
\includegraphics[height=5.4cm]{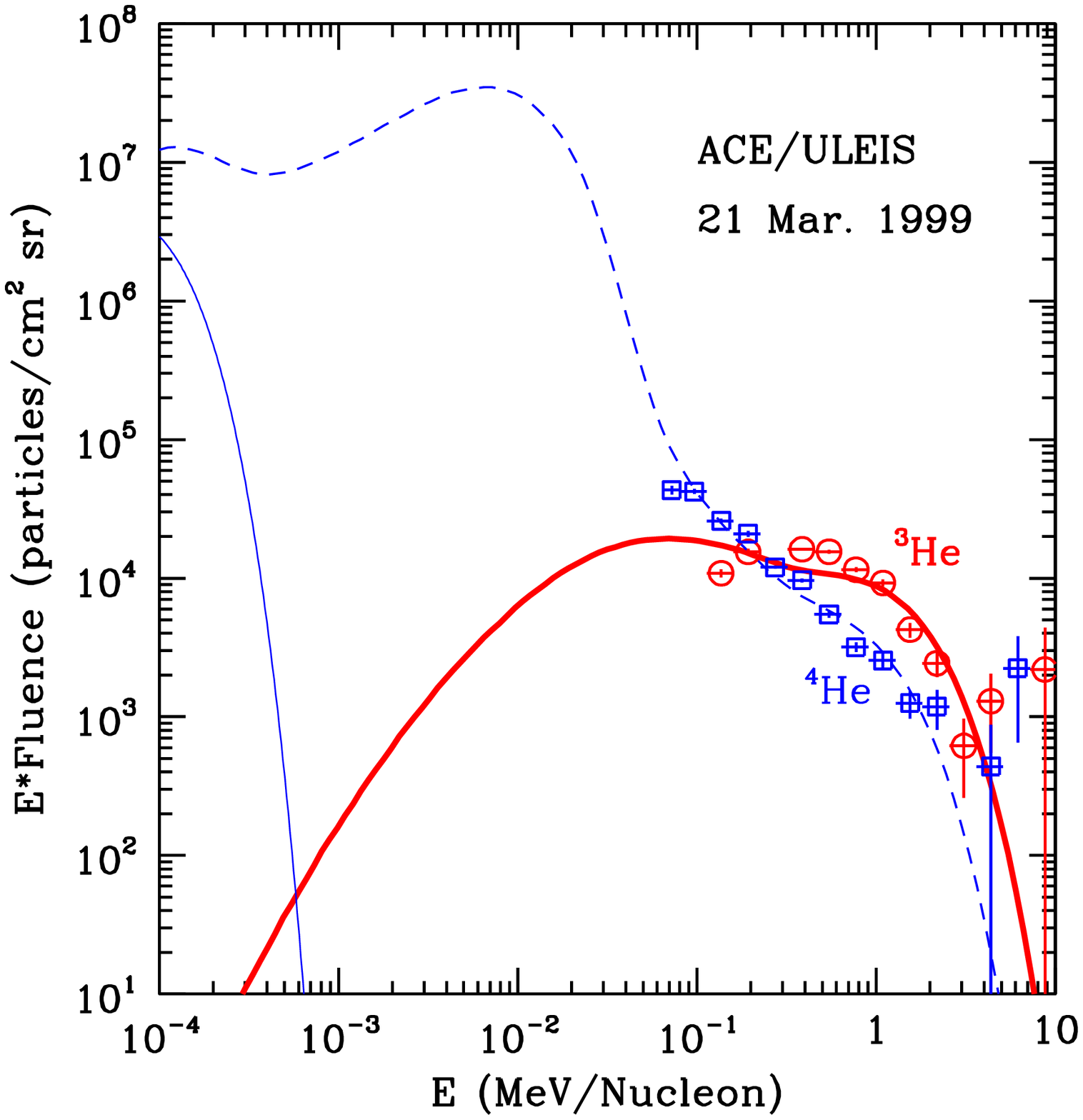}
\end{center}
\begin{center}
\includegraphics[height=5.4cm]{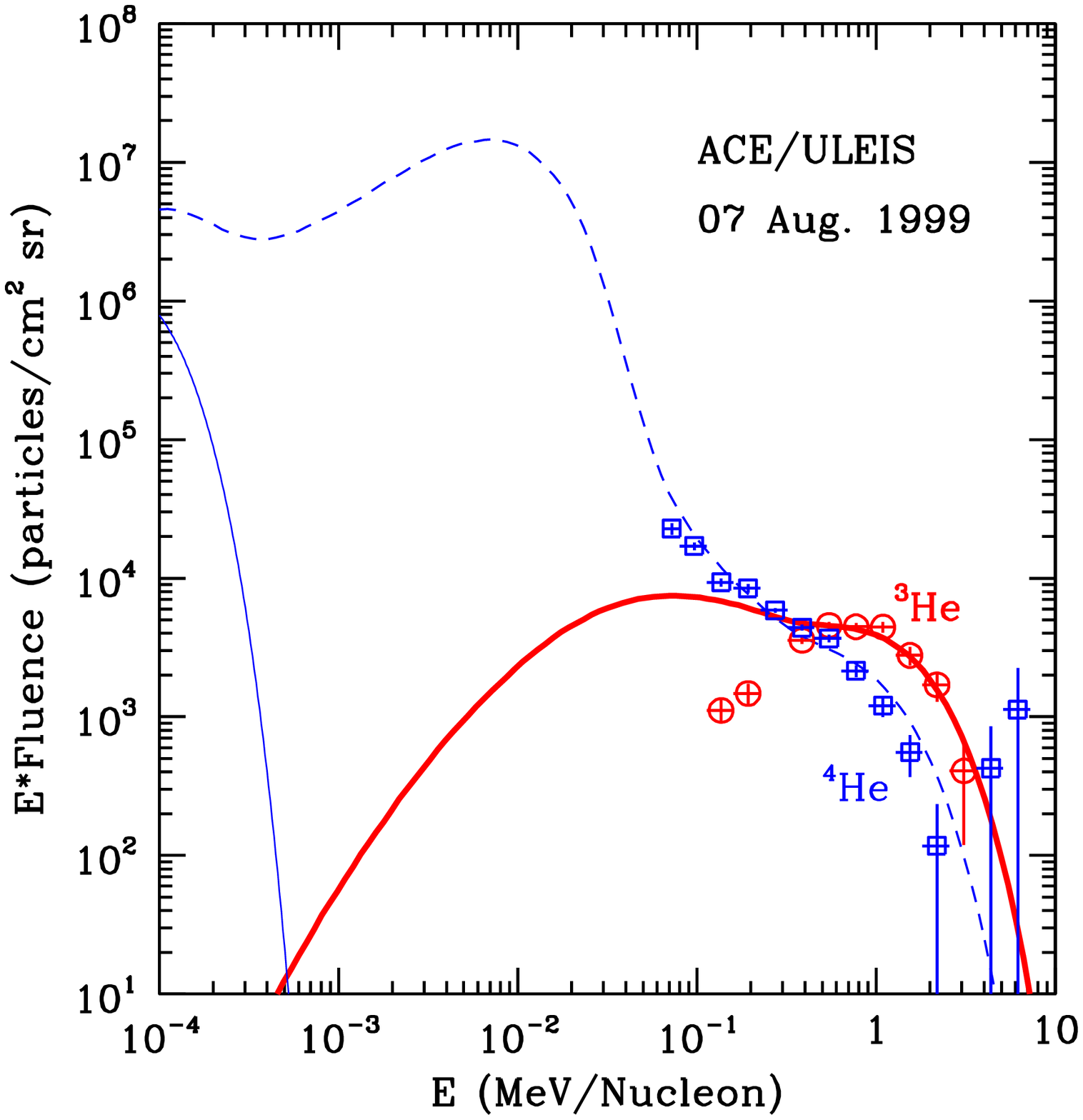}
\hspace{-0.6cm}
\includegraphics[height=5.4cm]{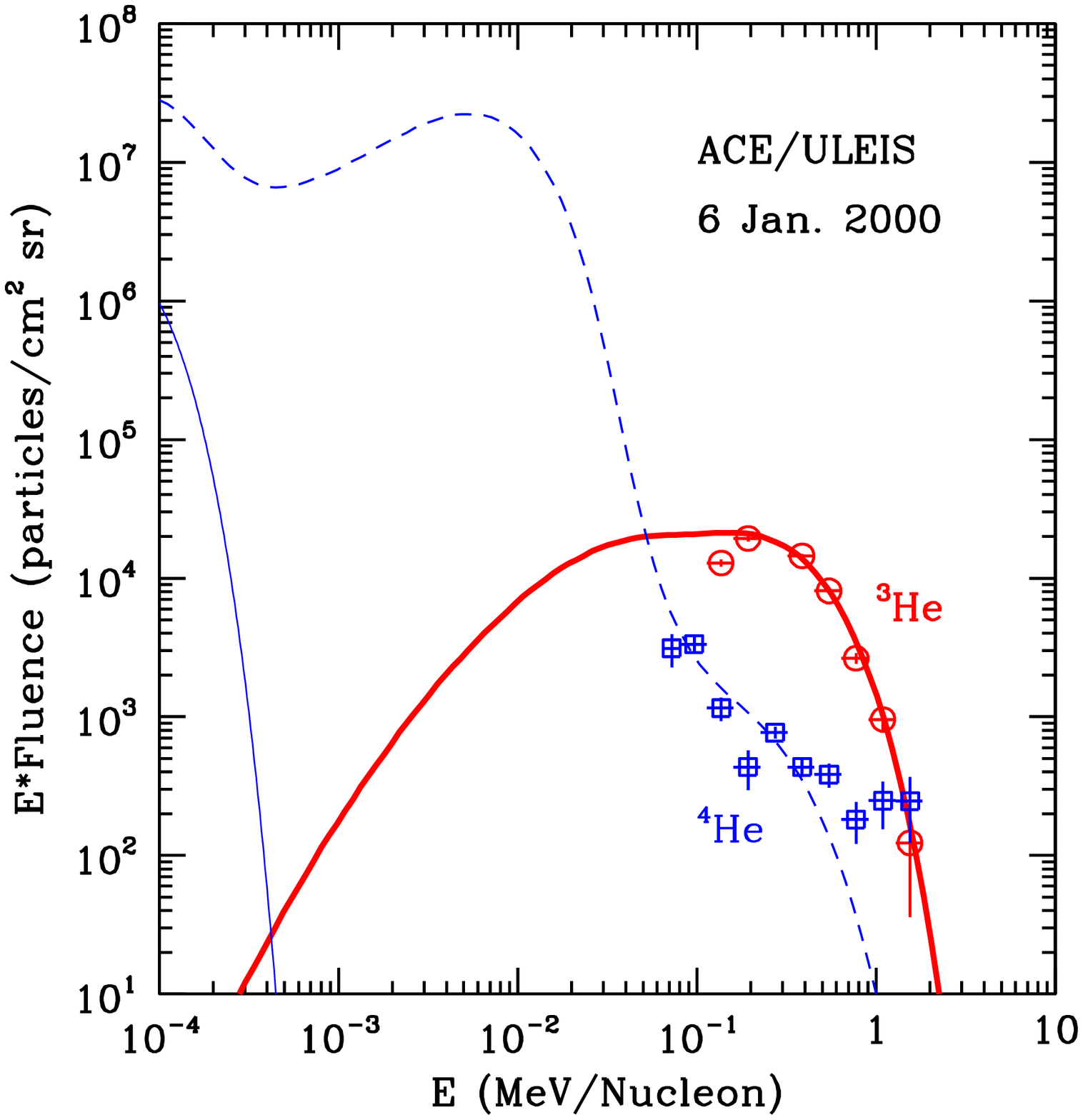}
\hspace{-0.6cm}
\includegraphics[height=5.4cm]{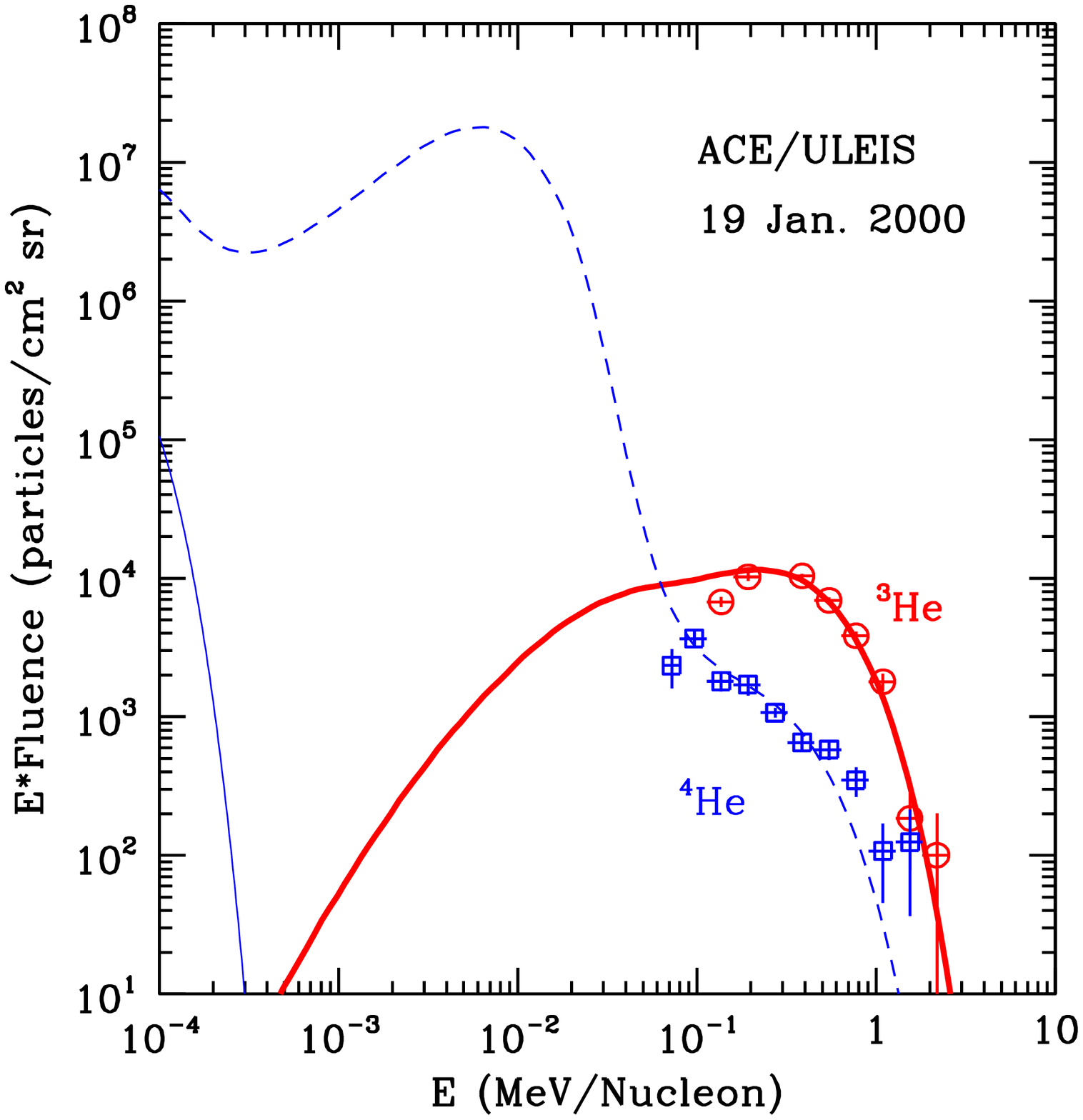}
\end{center}
\caption{
Same as the right panel of Figure \ref{fig7.ps}, but for five impulsive SEP events with convex $^3$He 
spectra observed by {\it 
ACE} between 1998 and 2000. The model parameters are listed in Table \ref{tab:mods}. See text 
for details.
}
\label{fig8.ps}
\end{figure}

The achievements of the model in explaining the 1999 September 30 event tempt us to apply it to
other impulsive SEPs.  We first consider the events with the so-called rounded spectra, as
suggested by the model calculations above (Fig. \ref{fig7.ps}).  Another five such impulsive
events are found during the {\it ACE} observation period from 1998 to 2000. These events show
quite different $^3$He enhancement, and the location of the high-energy cutoffs of the spectra
varies considerably (Figure \ref{fig8.ps}).  

In the previous section we have shown that the relative acceleration of $^3$He and $^4$He is
controlled by the thermal damping of waves at small scales since these damping effects
only modify the acceleration rate of $^4$He at low energies. In the application of the SA
model to the event studied above, we find that the acceleration of $^3$He is quite
insensitive to the temperature of the injected plasma because its acceleration time is always
much shorter than its loss time. However, the acceleration of $^4$He can be suppressed
dramatically as $T$ decreases because of the difficulty in the direct acceleration at low
energies. The three model parameters, $T$, $k_{\rm max}$, and $q_{\rm h}$, thus have the
equivalent function of controlling the acceleration of low-energy $^4$He, and consequently the 
enhancement of $^3$He. These three parameters
are all related to the thermal damping processes of plasma waves. For a given energy input rate
into the turbulence at large scales, the value of $k_{\rm max}$ and $q_{\rm h}$ should only depend on 
$T$. In what follows we assume that $k_{\rm max}$ and $q_{\rm h}$ weakly depend on the turbulence 
energy density and $T$ and fix them to the values used for the event shown in Figure 
\ref{fig7.ps}. The corresponding effective temperature of the injected plasma $T_{\rm eff}$ is 
used to adjust the enhancement of $^3$He (or more accurately the suppression of accelerated $^4$He).

The analyses of the wave-particle interactions in the previous section also suggest that the 
high-energy cutoff of the accelerated ion spectra is likely related to the
generation length of waves in the PC branch. We therefore leave $k_{\rm min}$ as another 
primary model parameter to account for the observed variations of the ion high-energy 
spectral cutoffs. For different solar flares $\tau_p$ and other model parameters will also change.  
We adjust $\tau_p$ and the normalization factor for the total amount of the injected particles 
$N_0\propto \dot{Q}$ to finalize the spectral fitting, but in order to reduce the amount of 
calculation we keep all other model parameters the same as those in the above model. 
Figure \ref{fig8.ps} shows the model fit to the spectra of these five events. The corresponding 
model parameters are list in table \ref{tab:mods}. The models in general give reasonable 
spectral fits to the five events, and reproduce their varied degree of $^3$He enhancement. 
There are, however, obvious discrepancies 
in the low-energy $^3$He spectrum of the 1999 August 7 event and the high-energy $^4$He 
spectra of the 2000 events.

\begin{table}[htb]
\begin{center}
\caption{Events and the Corresponding Model Parameters.
\label{tab:mods}}
\vskip 0.2in
\begin{tabular}{lcccc}
\tableline\tableline
Events   & $k_{\rm min}/k_{\rm max}$ & $\tau_{p}^{-1}/10^{-3}$s$^{-1}$ &
$k_{\rm B}T_{\rm eff}$/keV  & $N_0$\tablenotemark{\dag} \\
\tableline
1998 Aug 18 & 0.16 & 6.0 & 0.12 & 0.79
\\
1999 Mar 21 & 0.075 & 6.0 & 0.14 & 1.6
\\
1999 Aug 7 & 0.075 & 6.5 & 0.12 & 0.63 
\\
1999 Sep 30 & 0.10 & 5.5 & 0.26 & 1
\\
2000 Jan 6 & 0.14 & 5.5 & 0.10 & 1.6 
\\
2000 Jan 19 & 0.14 & 7.5 & 0.07 & 0.79 
\\
\tableline
\end{tabular}
\tablenotetext{\dag}{The normalization is with respect to the event on 1999 September 30.
}
\tablecomments{
The other model parameters are the same as those for the 1999 September 30 event, i.e. $L = 
2\times 10^9$ cm, $B_0 = 200$ G, $n_e = 9\times 10^8$ cm$^{-3}$ (therefore $\alpha = 
0.48$), $k_{\rm max} = 2\alpha\delta^{-1/2}c\Omega_p^{-1}$ for the PC branch and one half of it 
for the HeC branch, $q_{\rm l} = 2$, $q = 2$, and $q_{\rm h} = 4$. }
\end{center}
\end{table}

For the six events studied here the effective temperature of the injected plasma varies 
within a factor of 4. Because the $^3$He enhancement decreases with the increase of $T_{\rm 
eff}$ (see discussions above and more quantitative studies in the next section), and the 
thermal damping effects are enhanced with temperature, we expect more 
significant variations in the real temperatures of the background plasmas for these events 
except that there is a positive correlation between the temperature $T$ and the turbulence 
energy density (indicated by the value of $\tau_p^{-1}$, which is anti-correlated with the 
$^3$He enhancement, because $^3$He can be accelerated to high energies even with large $\tau_p$ 
while $^4$He can not). The ``generation'' length scale or $k_{\rm min}$ of the waves in the PC branch 
changes within a factor of 2.2, which seems to be a reasonable range.  The characteristic timescale 
$\tau_p$ has much less variations (about 30\%), which may be due to the similarities of the 
rounded spectra of these events and the comparable amplitude of the events as indicated by the 
relative normalization factor in the last column, which varies within a factor of 3.

\clearpage

\subsection{Exploring of Model Parameters Space}
\label{dis}

In the previous section we have shown that the observed degree of enhancement and the primary 
features of the rounded spectra of $^3$He and $^4$He can be reproduced in our model with reasonable 
values of the parameters. Here we describe the dependence of these observed characteristics on 
various parameters.

\subsubsection{Turbulence Spectrum and Thermal Damping}

\begin{figure}[htb]
\begin{center}
\includegraphics[height=5.4cm]{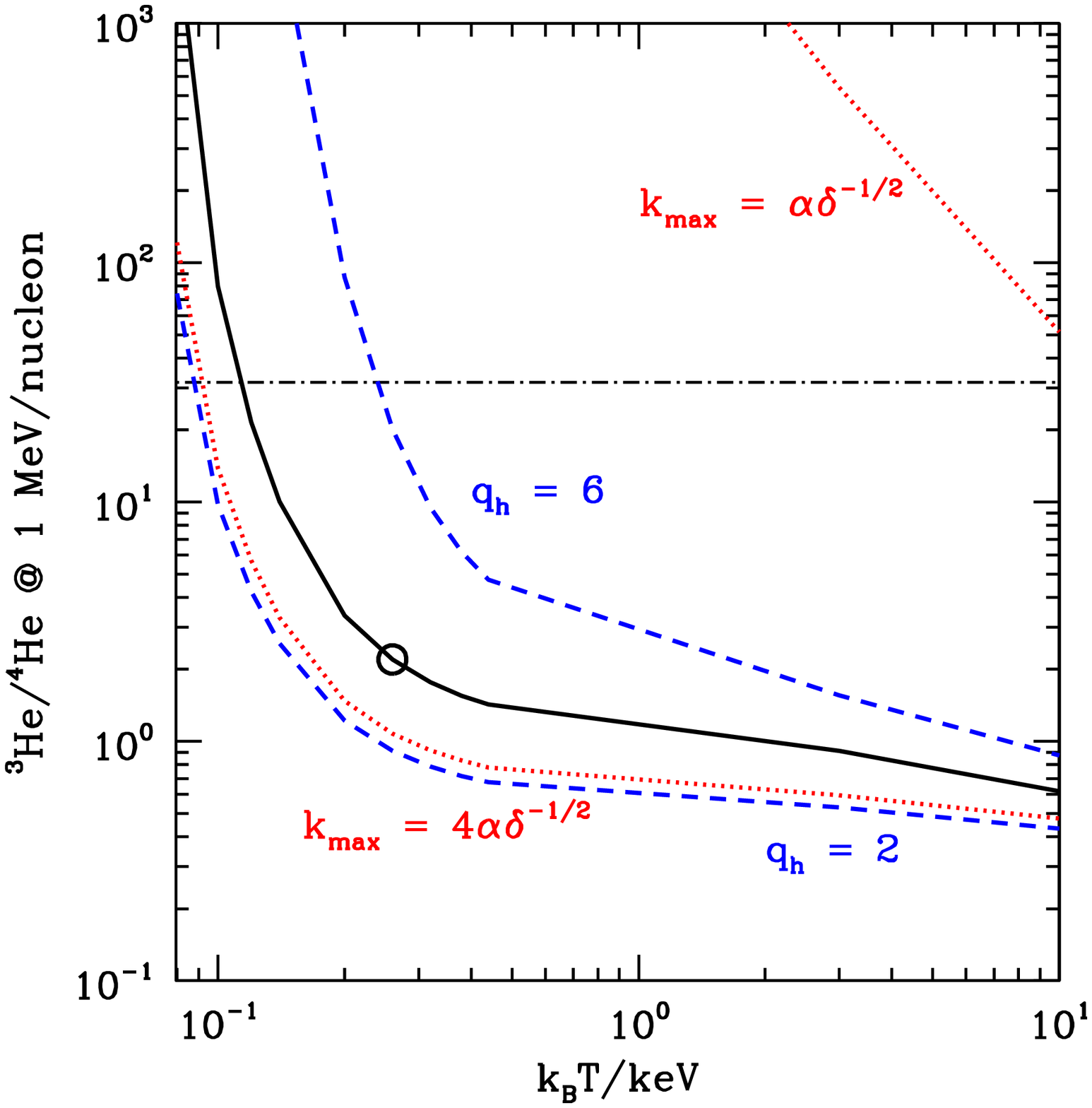}
\hspace{-0.6cm}
\includegraphics[height=5.4cm]{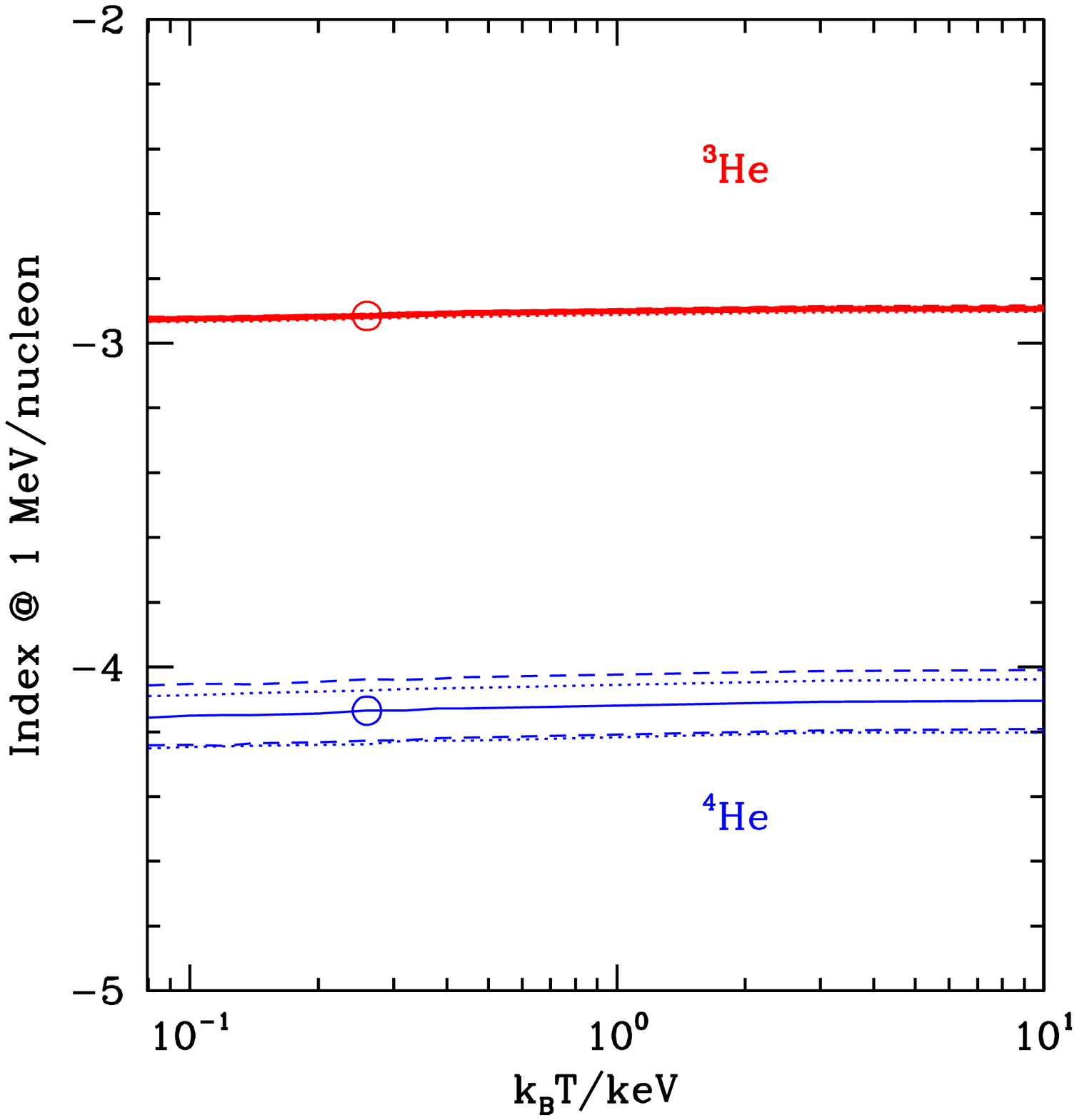}
\hspace{-0.6cm}
\includegraphics[height=5.4cm]{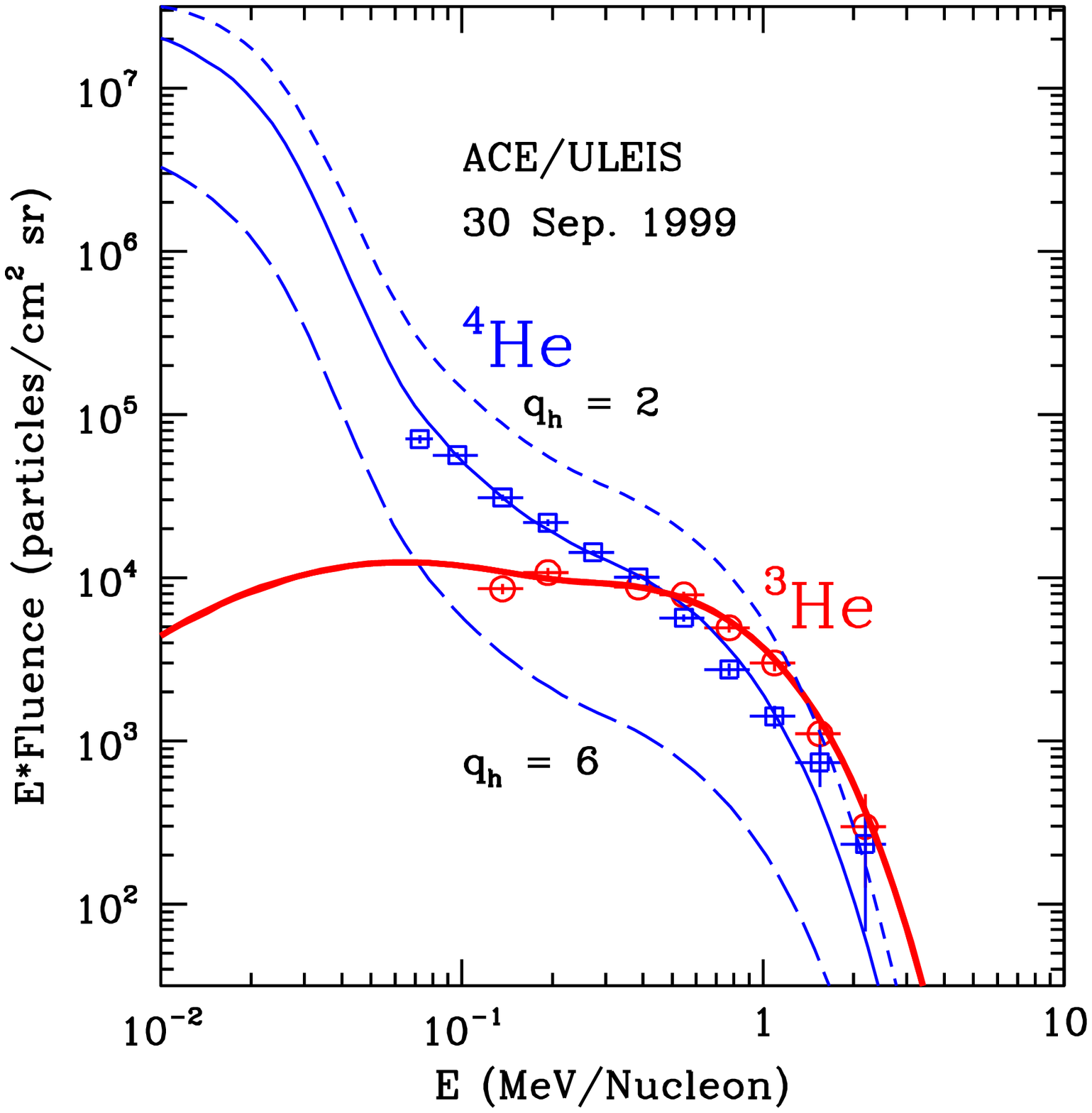}
\end{center}
\caption{
{\it Left:} Dependence of the $^3$He to $^4$He ratio at 1 MeV/nucleon on the temperature of the 
injected plasma for several models with different turbulence spectra. The solid line 
corresponds to the fiducial model for the 1999 September 30 event, which is indicated by 
the large open circle. The dotted lines are for two models with $k_{\rm max} = 
\alpha\delta^{-1/2}$ ({\it upper one}) and $4\alpha\delta^{-1/2}$ 
({\it lower one}). The other model parameters are the same as those for the 
solid line. The same is true for the dashed lines except now $q_{\rm h} = 6$ ({\it upper 
one}) and 2 ({\it lower one}). The latter has no thermal damping. The horizontal 
dot-dashed line gives the highest $^3$He enhancement observed so far, corresponding to the 
event on 2000 January 6. Most of the models with strong damping ($k_{\rm max} = \alpha\delta^{-1/2}$) 
or very low temperatures ($<10^6$ K) can be ruled out because the very high predicted $^3$He 
enhancement has never been observed. {\it Middle:} Same as the left panel except for the spectral 
indexes of $^3$He 
({\it thick lines}) and $^4$He ({\it thin lines}) defined as ${\rm d}\ln{f}/{\rm d}\ln{E}$ at 1 
MeV/nucleon. {\it Right:} The dependence of the accelerated $^3$He and $^4$He spectra on $q_{\rm 
h}$. The upper dotted line is for $^4$He and $q_{\rm h} = 2$ and corresponds to the case 
without thermal damping. The lower dotted line is for $q_{\rm h} = 6$. The solid and dashed 
lines and the data points correspond to the fiducial model shown in the right panel of Figure 
\ref{fig7.ps}. All other model parameters remain the same. The $^3$He acceleration is not 
affected by $q_{\rm h}$.
} \label{fig9.ps}
\end{figure}

One of the most uncertain part of the model is related to the characteristics of turbulence. The 
two model parameters $k_{\rm max}$ and $q_{\rm h}$ characterize the wave damping process. The 
background temperature which plays important roles in this damping process also has important 
influence. As mentioned above, thermal damping that involves complicated nonlinear processes 
determines the connections among these parameters (Yan, Petrosian, \& Lazarian 2005). Investigation 
of these aspects is beyond the scope 
of this paper. Here we consider the influence of these parameters separately. In Figure \ref{fig9.ps} 
we show the effects of the above mentioned parameters, while keeping other parameters constant and 
equal to those in Figure \ref{fig7.ps}. The enhancement of $^3$He (or equivalently the suppression of 
$^4$He acceleration) decreases rapidly with increasing $T$ at low values of $T$, primarily because 
the number of $^4$He in the energy range where $\tau_{\rm loss}<\tau_{\rm a}$ decreases 
exponentially with T. However, even with a 
high temperature, the acceleration of $^4$He  is still suppressed because its escape process 
dominates in the tens of keV energy range, and the $^3$He enhancement eventually saturates. The 
saturation level depends on the other parameters. For higher values of 
$k_{\rm max}$ and lower $q_{\rm h}$ (indicating slower damping) the enhancement of $^3$He is smaller. 
But for the range of values explored here, there is always a significant enhancement. Even in the 
case without thermal damping effects, i.e. for $q_{\rm h}=2$, strong $^3$He enhancement can still be 
produced, which underscores the importance of using the exact dispersion relation in modeling 
the wave-particle interactions (LPM04).  The thermal damping affects primarily the $^3$He 
to $^4$He ratio, but introduces minor changes in the spectrum of accelerated ions (specially 
$^3$He) ({\it middle and right} panels).  

The dot-dashed horizontal line in Figure \ref{fig9.ps} indicates the observed range of $^3$He
enhancement. The spectral indexes vary from -1 to -4. The lack of observations with even higher 
$^3$He to $^4$He ratio speaks against models with small $k_{\rm max}$ (or very strong damping) and 
low temperatures of the background
plasmas. More detailed theoretical investigations of the wave damping should be able to resolve
this issue and give better constraints on the SA model.

\clearpage

\subsubsection{Turbulence Strength and Scale}

\begin{figure}[htb]
\begin{center}
\includegraphics[height=5.4cm]{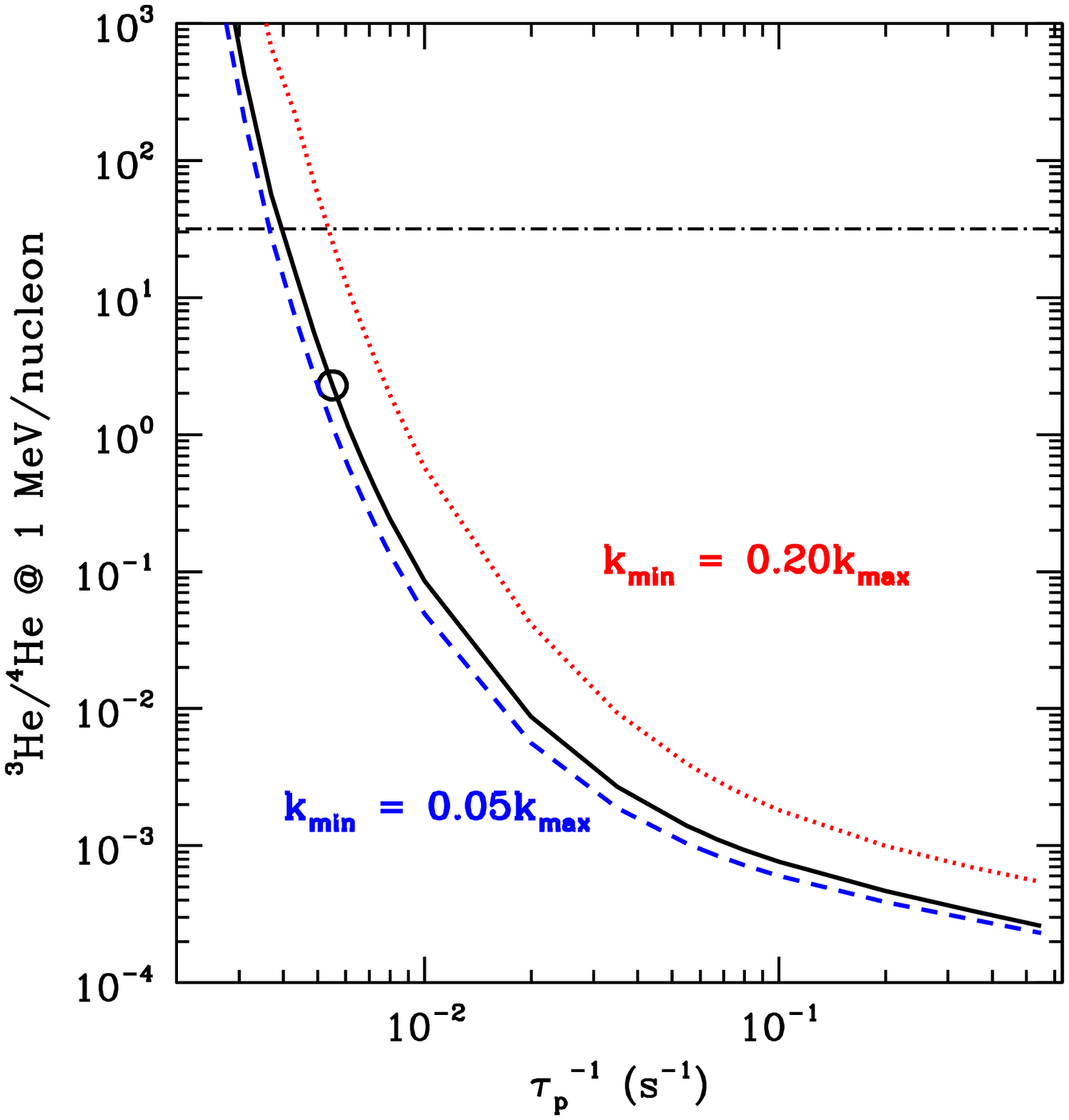}
\hspace{-0.6cm}
\includegraphics[height=5.4cm]{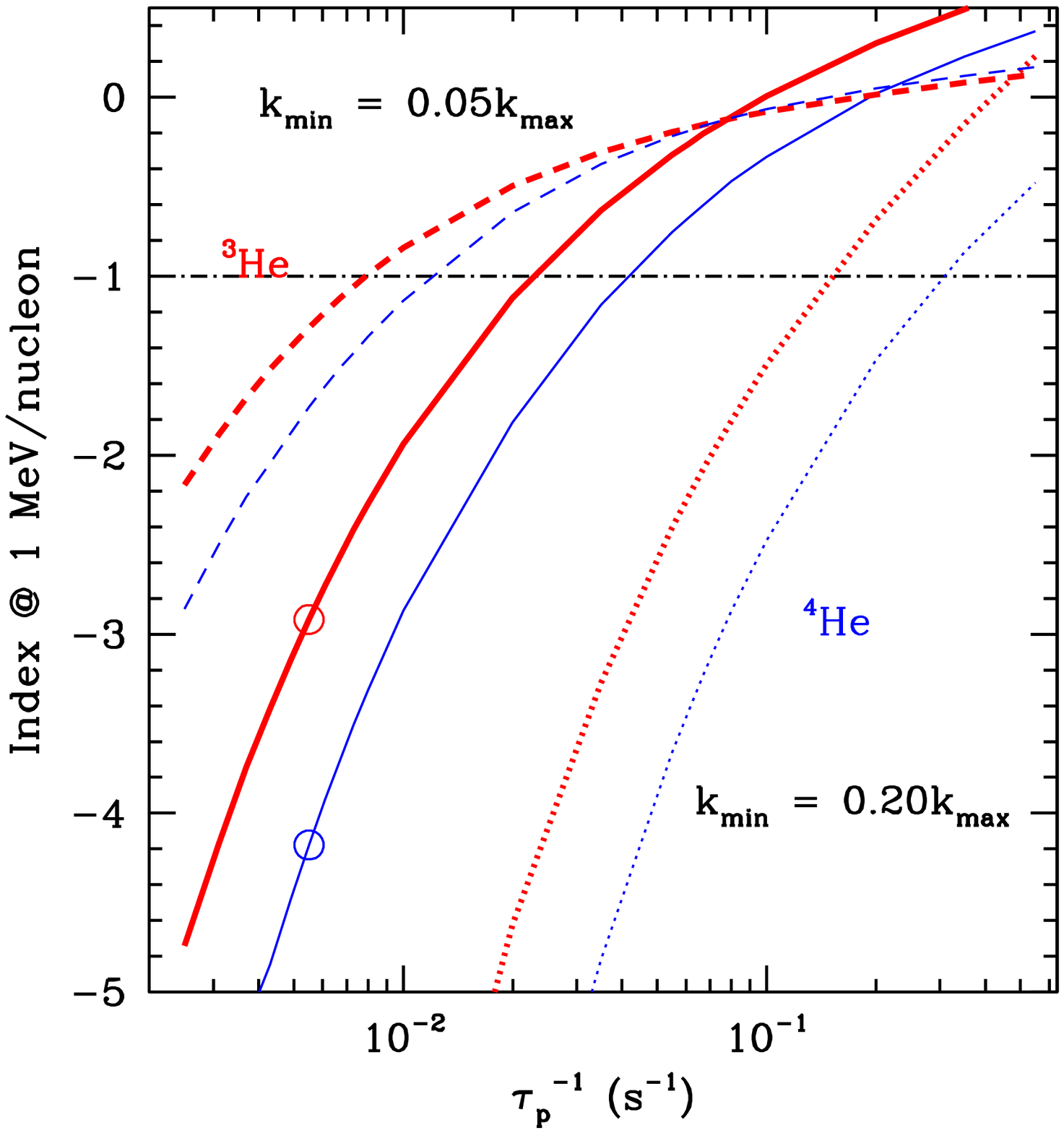}
\hspace{-0.6cm}
\includegraphics[height=5.4cm]{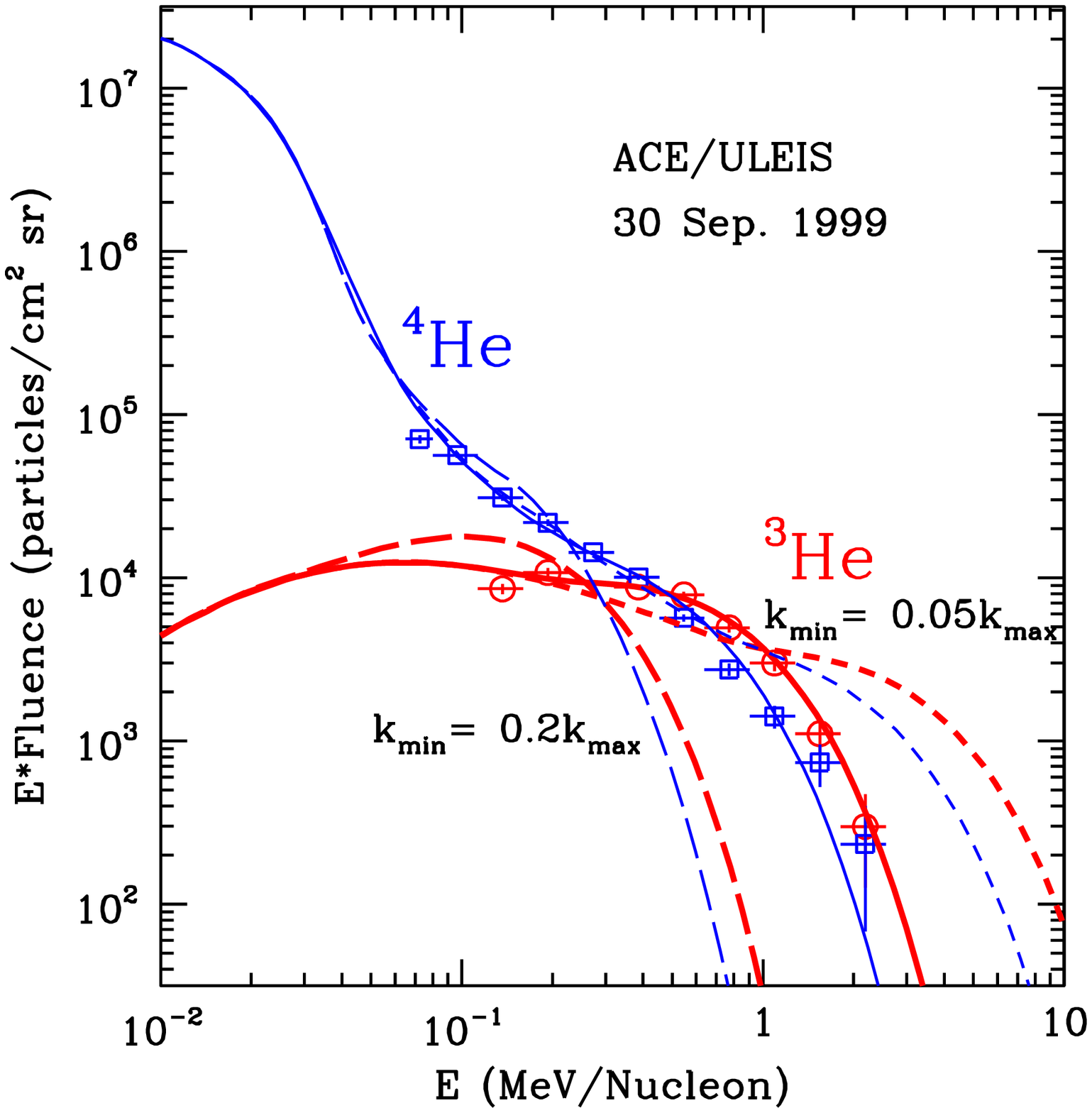}
\end{center}
\caption{
Same as Figure \ref{fig9.ps}. {\it Left:} for the dependence on $\tau_p^{-1}$ and for two dynamic
ranges $k_{\rm min}/k_{\rm max} = 0.2$ ({\it upper dotted line}) and $0.05$ ({\it lower dashed line}) 
with a fixed $k_{\rm max} = 2 \alpha\delta^{-1/2}$. Events with weak 
turbulence have high $^3$He enhancement and soft spectra. With very strong turbulence 
($\tau_p^{-1}\ge0.1$), all of the injected $^3$He and $^4$He are 
accelerated to high energies, and $^3$He enhancement disappears. {\it Middle:} Here the dashed 
lines are for $k_{\rm min} = 0.1 \alpha\delta^{-1/2}$ and the dotted lines are for 
$k_{\rm min} = 0.4 \alpha\delta^{-1/2}$. 
{\it Right:} The dependence of spectra on $k_{\rm min}$ showing that for lower $k_{\rm min}$ the 
spectra cut off at higher energies. The thinner lines are for $^4$He and the thicker for 
$^3$He.
}
\label{fig10.ps}
\end{figure}

The ``generation'' length scale of waves in the PC branch, characterized by $k_{\rm min}$, and the 
turbulence intensity, proportional to $\tau_p^{-1}$, are two other important parameters characterizing 
the PWT. Figure \ref{fig10.ps} shows the dependence of the $^3$He enhancement and the accelerated 
spectra of $^3$He and $^4$He on these parameters. In general, again almost all of the injected 
$^3$He ions are accelerated to high energies, but the acceleration of low-energy $^4$He ions is very 
sensitive to $\tau_p^{-1}$ specially at low $\tau_p^{-1}$. Therefore, the $^3$He to $^4$He ratio 
at high energies decreases rapidly with the increase of the turbulence intensity. With very strong 
turbulence ($\tau_p^{-1}\ge 0.1$), almost all of the injected $\alpha$-particles could be accelerated 
to high energies, and there is essentially no $^3$He enhancement (see {\it left} panel). 
The high-energy spectra of the accelerated particles are also
very sensitive to $\tau_p^{-1}$ and $k_{\rm min}$.  With the increase of $\tau_p^{-1}$, the spectra
become harder ({\it middle} panel). But for low values of $k_{\rm min}$, the cutoff energies of the 
accelerated ions will increase with minimal effects on the shapes of the spectra at lower energies 
({\it right} panel). 

\subsubsection{Background Plasma Parameters}

\begin{figure}[htb]
\begin{center}
\includegraphics[height=5.4cm]{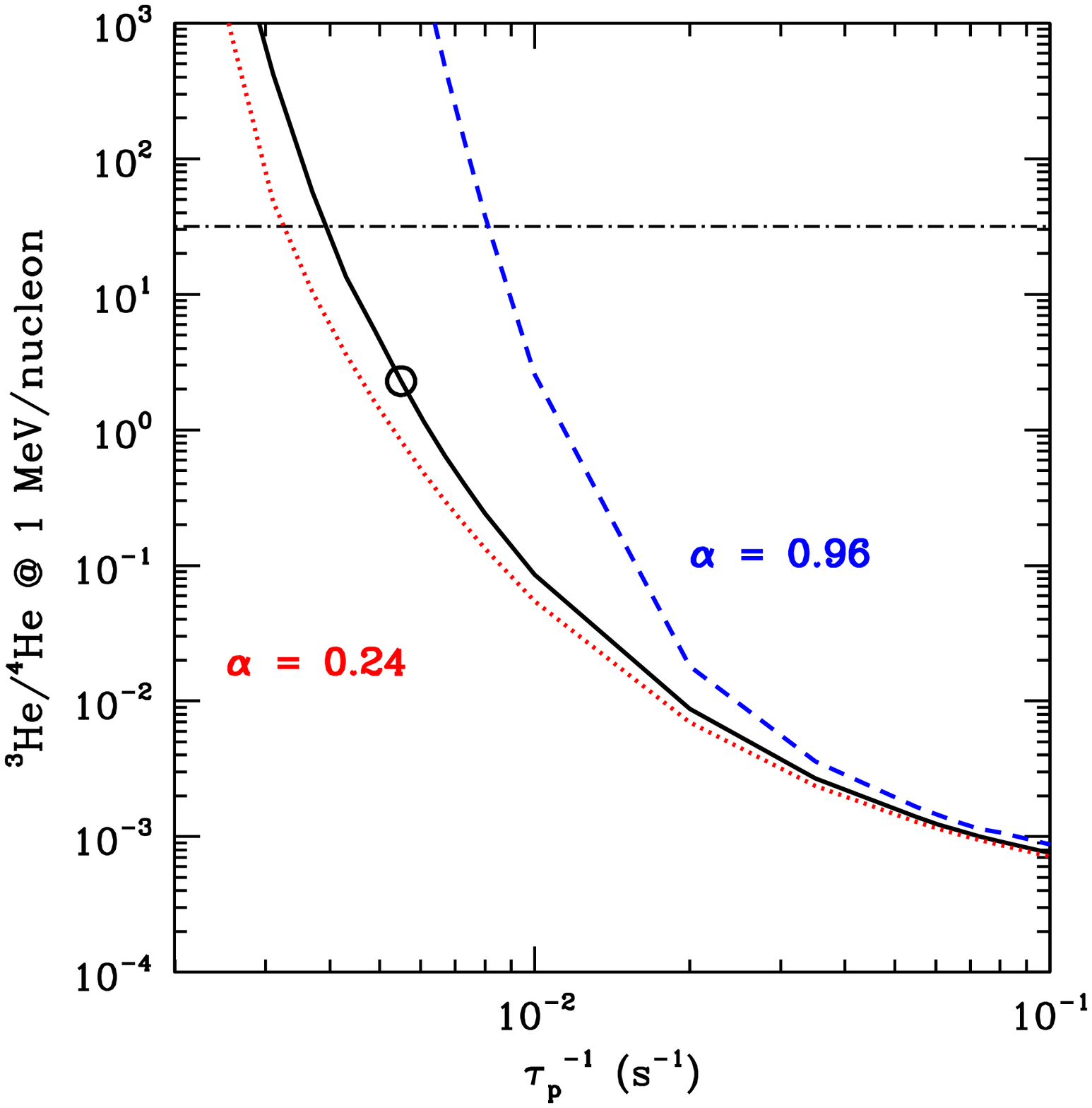}
\hspace{-0.6cm}
\includegraphics[height=5.4cm]{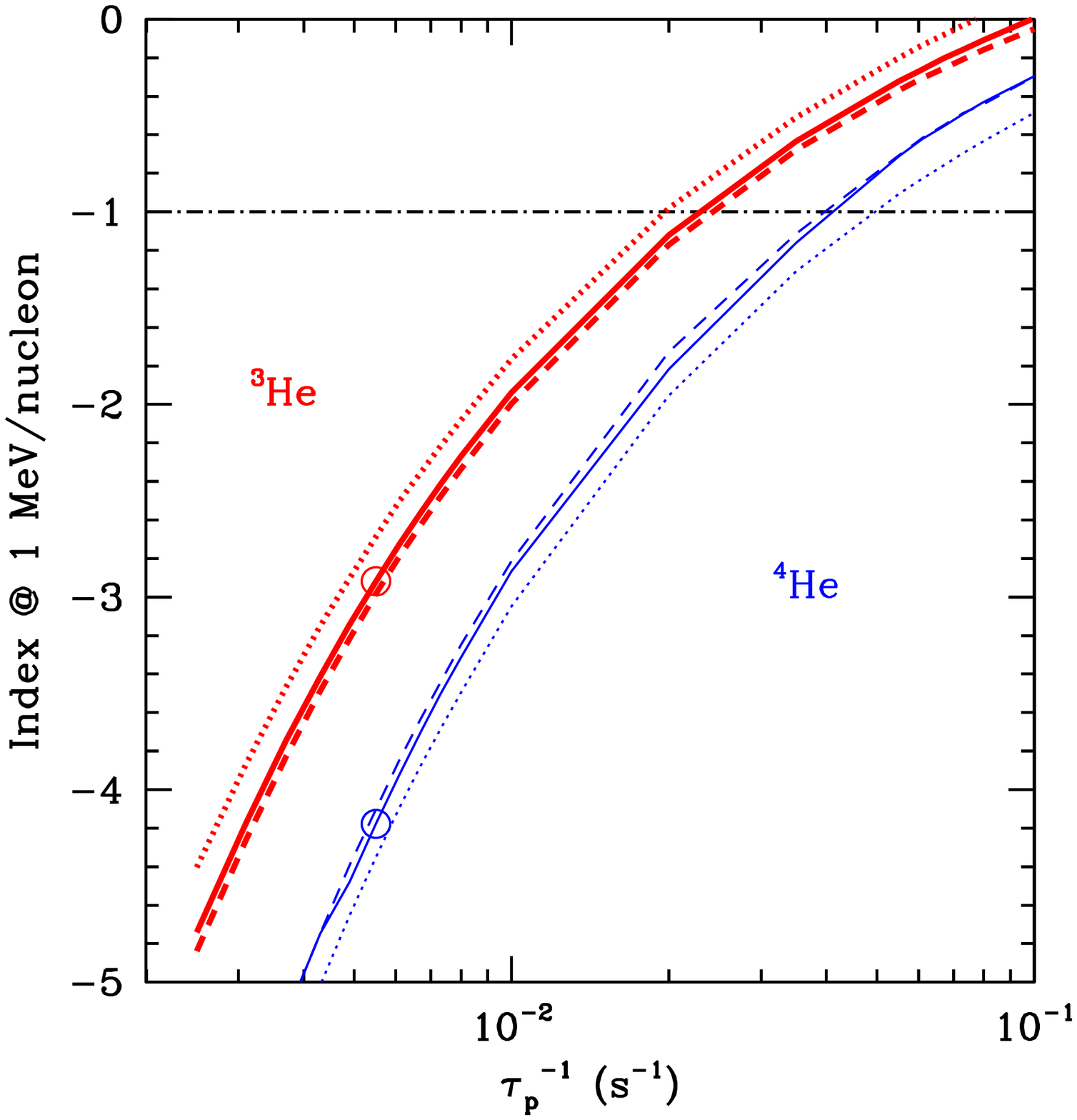}
\hspace{-0.6cm}
\includegraphics[height=5.4cm]{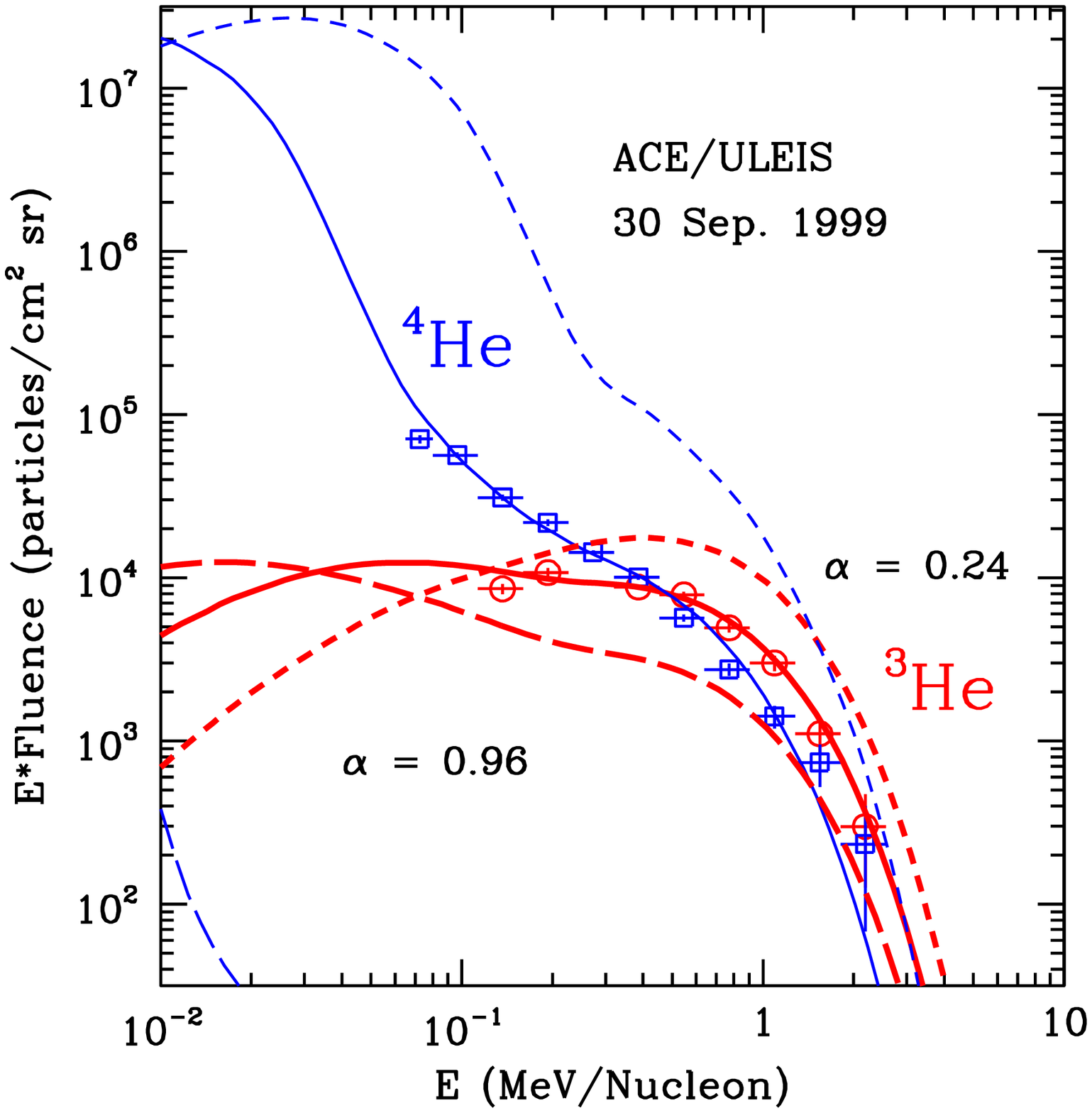}
\end{center}
\caption{
Same as Figure \ref{fig10.ps}, but for models with different plasma parameter $\alpha$. Here 
the dashed and dotted lines in the left and middle panels are for $\alpha = 0.96$ and 0.24, 
respectively. Different values of $\alpha$ are achieved by adjusting the large scale magnetic 
field $B_0$, which affects the particle acceleration only via $\alpha$.  The gas density $n$ 
remains the same so that the loss and escape times do not change.  All other model 
parameters and $k_{\rm max}/\alpha$ also remain the same.  Models with larger 
$\alpha$ have less efficient particle acceleration at low energies, giving rise to softer 
low-energy spectra ({\it right}). (Note that because the $^4$He acceleration is very 
inefficient in the case of $\alpha = 0.96$, the spectrum of the accelerated $\alpha$-particle 
is barely seen at the low left corner in the figure.) The high-energy spectral shape is not 
very sensitive to $\alpha$ ({\it middle}).
}
\label{fig11.ps}
\end{figure}

Besides the temperature, other characteristics of the background plasma also affect the particle 
acceleration, with the main parameter being the plasma parameter $\alpha$.  To change $\alpha$ we 
keep the density, which affects the loss and escape processes at low energies, constant and change 
the large scale magnetic field $B_0$.  Figure \ref{fig11.ps} depicts the influence of $\alpha$. 
Following the right panel of Figure \ref{fig5.ps}, we keep $k_{\rm max}/\alpha$ and all other model
parameters unchanged.  As expected, $\alpha$ predominately affects the ion acceleration in
the low-energy range.  The $^3$He enhancement decreases with the decrease of $\alpha$, 
corresponding to more strongly magnetized plasmas. The $\tau_p$ dependence of the $^3$He to 
$^4$He ratio at 1 MeV/nucleon is quite different for different values of $\alpha$, especially 
when $\tau_p^{-1}$ is small ({\it left} panel). The high-energy spectral indexes, however, are 
only weakly dependent on $\alpha$ ({\it middle} panel). The right panel shows explicitly that 
the acceleration of low-energy ions, specially $^4$He, becomes more efficient in more strongly 
magnetized 
plasmas.

\section{SUMMARY AND CONCLUSION}
\label{sum}

Recent observations of SEPs and radiations produced by solar flares indicate that SA may play a major
important role in the acceleration of all particles. It has been applied
extensively to impulsive SEPs, and may be involved in the pre-acceleration of particles in gradual 
flares, where the particles are further accelerated to even higher energies by interplanetary shocks 
(Desai et al. 2001, 2003, 2004).  We have described a SA model for the acceleration of $^3$He and 
$^4$He in SEPs by PWT assuming some characteristics for this turbulence. We have shown that for 
reasonable values of parameters describing the PWT and the background plasma the model gives 
acceptable fit to the spectra and explains the extreme enhancement of $^3$He. Model fits are obtained 
for a few events. In general relative to the loss rate $^3$He acceleration rate is much higher than 
that of $^4$He. All $^3$He particles are accelerated while only a small fraction of $^4$He ions 
attain higher energies. The primary reason for this is that the presence of a significant 
abundance (relative to protons) of $\alpha$-particles in the background plasma allows excitation of 
PC branch waves. It is shown that nonrelativistic $^3$He 
and $^4$He ions resonate mostly with waves with frequencies close to the $\alpha$-particle 
gyro-frequency. To study the SA of these ions the exact dispersion relation for the plasma waves must 
be used, resulting in more efficient acceleration than scattering that could lead to anisotropic 
particle distributions. The presence of cosmic abundance of $\alpha$-particles in the background 
plasma then play a key role in determining the relative acceleration of the two ions, because of
their modification to the dispersion relation, so that low-energy $^4$He ions interact only with low 
phase velocity waves giving 
rise to a long acceleration time. In addition, the stronger the damping of the waves, (i.e., the 
lower the value of the cutoff $k_{\rm max}$ and/or the larger the cutoff index $q_{\rm h}$), the 
stronger the 
suppression of the acceleration of $^4$He and the higher the enhancement of $^3$He. On the other 
hand, a higher background temperature (which could cause a stronger damping) increases the efficiency 
of $^4$He acceleration up to tens of keV. 

$\bullet$ The turbulence intensity determines the hardness of the accelerated particle spectra; 
Harder spectra and smaller $^3$He enhancements arise from stronger turbulence.

$\bullet$ In a strongly magnetized plasma (lower value of $\alpha$) the condition becomes more 
favorable for acceleration of $^4$He and the $^3$He enhancement decreases. The $\alpha$ parameter 
also affects the ion spectra.

$\bullet$ The effect of the lower wavenumber $k_{\rm min}$ (large scale) cutoff in the spectrum of 
turbulence is 
felt at high energies by both ions; the cutoff energy increases for lower values of $k_{\rm min}$.


We note, however, that not all feature of observed spectra can be described easily by the model.
In the MeV/nucleon energy range both $^3$He and $^4$He ions interact mostly with the same
waves in the PC branch, the acceleration time of $^4$He is always longer than that of $^3$He.
As a result, the model predicts a$^3$He spectrum that is always harder than that of $^4$He. This is 
in conflict with the observed decrease of $^3$He to $^4$He ratio with energy above a few
MeV/nucleon in some flares (Mason et al. 2002; Reames et al. 1997; M\"{o}bius et al.  1982;  
M\"{o}bius et al. 1980). This may suggest inadequacy of the simple model used here; e.g. it is 
possible that
at high energies obliquely propagating waves becomes more important.
Or this may requires a special injection process or a second-phase shock acceleration
(Van Hollebeke, McDonald \& Meyer 1990; Serlemitsos \& Balasubrahmanyan 1975; Reames et al.
1997).  

In the model described here the PC branch plays a dominant role in the acceleration of $^3$He and 
$^4$He ions. Detailed studies of the generation of waves in this branch and a comprehensive 
investigation of 
the thermal damping effects will reduce the model parameters considerably, giving better constraints 
on the SA models. It is clear that the model can also be used to study the acceleration of 
heavy-ions. The heavy-ion acceleration is dominated by resonant interactions with waves in the HeC 
branch, and ions with lower charge-to-mass ratio are accelerated more efficiently due to their
interactions with larger scale waves (Mason et al. 1986; Mason et al. 2004; Reames \& Ng
2004). With simultaneous observations of solar flares by the {\it WIND} and {\it ACE} 
spacecrafts, a comprehensive study of acceleration of all charged particles including 
electrons, protons, and ions by the same PWT will set strict constraints on the power spectrum 
of PWT, guiding the exploration of the dynamical properties of the plasma turbulence.  A 
complete time dependent treatment of the wave-particle interaction, including the coupled 
kinetic evolution of the particle distribution and the turbulence spectrum is also required.

\acknowledgments

This research was partially supported by NSF grant ATM-0312344, NASA grants NAG5-12111, 
NAG5 11918-1 at Stanford and NASA grant PC 251429 at University of Maryland.

\end{document}